\documentclass[fleqn,usenatbib]{mnras}
\usepackage{newtxtext,newtxmath}
\usepackage[T1]{fontenc}
\usepackage{pdfsync}
\DeclareRobustCommand{\VAN}[3]{#2}
\let\VANthebibliography\thebibliography
\def\thebibliography{\DeclareRobustCommand{\VAN}[3]{##3}\VANthebibliography}

\usepackage{graphicx}	% Including figure files
\usepackage{amsmath}	% Advanced maths commands

\graphicspath{{figures/}}

\usepackage{xcolor}
\definecolor{c-enlace}{HTML}{2854d7}
\usepackage{hyperref}
\hypersetup{breaklinks=true, colorlinks = true, urlcolor = c-enlace, linkcolor = c-enlace, citecolor = c-enlace}

\title[Thermal forces on embedded embryos]{On the interaction of pebble accreting embryos with the gaseous disc: importance of thermal forces}

\author[Cornejo et al.]{
Sonia Cornejo,$^{1}$\thanks{E-mail: soniac@icf.unam.mx}
Fr{\'e}d{\'e}ric S. Masset,$^{1, 2}$
and F. J. S{\'a}nchez-Salcedo$^{3}$
\\
$^{1}$Instituto de Ciencias F{\'i}sicas, Universidad Nacional Aut{\'o}noma de M{\'e}xico, Av. Universidad s/n, 62210 Cuernavaca, Mor., M{\'e}xico\\
$^{2}$University Nice-Sophia Antipolis, CNRS, Observatoire de la C{\^o}te d’Azur, Laboratoire LAGRANGE, CS 34229. 06304 Nice Cedex 4, France\\
$^{3}$Instituto de Astronom{\'\i}a, Universidad Nacional Aut{\'o}noma de M{\'e}xico, Apartado Postal 70-264, 04510 Mexico City, Mexico
}

\date{Accepted XXX. Received YYY; in original form ZZZ}

\pubyear{2023}

\begin{document}
\label{firstpage}
\pagerange{\pageref{firstpage}--\pageref{lastpage}}
\maketitle

\begin{abstract}
A planetary embryo embedded in a gaseous disc can grow by pebble accretion while subjected to a gravitational force from the disc that changes its orbital elements. Usually, that force is considered to arise from the Lindblad and corotation resonances with the embryo.  However, more important contributions exist for low-mass planets. Radiative thermal diffusion in the vicinity of embryos yields an additional contribution to the disc's force that damps the eccentricity and inclination much more vigorously than the resonant interaction with the disc, and that in general induces fast inward migration. In addition, the irradiation of the disc by a hot embryo gives rise to an additional contribution that excites eccentricity and inclination, and induces outward migration. Which of the two contributions dominates depends on the embryo's luminosity. We assess the importance of these contributions (termed thermal forces) on the dynamics and growth of a set of pebble-accreting embryos initially of Martian mass, by means of N-body simulations that include analytic expressions for the disc's force. We find very different outcomes for the embryos subjected to thermal forces and those subjected only to resonant forces. Importantly, we find that the median final mass of the embryos subjected to thermal forces is nearly independent of the metallicity, whereas this mass roughly scales with the metallicity when they are subjected only to resonant forces. These results can be explained by the strong damping of eccentricity and inclination at low metallicity, which enhances the embryos' accretion efficiency.
\end{abstract}


\section{Introduction}
\label{sec:introduction}
Planetary systems are formed in the discs of gas and dust that surround newly formed stars. Dust settles toward the midplane of the disc and sticks together to form larger objects (mm-cm particles) known as pebbles \citep{1977MNRAS.180...57W, Guttler2010, Zsom2010, Blum2018}. The accretion sequence extends by directly forming kilometre-sized planetesimals through hydrodynamic processes (e.g., the streaming instability) that concentrate solids into dense filaments that undergo gravitational collapse \citep{2005ApJ...620..459Y, Johansen2007}. These planetesimals remain embedded within a population of pebbles that did not participate in the collapse.

The formation of larger objects, called planetary embryos, occurs through two processes that can act simultaneously. One of these processes, the collisions of pairwise planetesimals, occurs first through runaway growth  \citep{Greenberg1978, Wetherill1993, 2000Icar..143...15K} followed by slower oligarchic growth \citep{2000Icar..143...15K,Chambers2006}. The other process, pebble accretion, happens through the aerodynamically enhanced capture of pebbles \citep{2010A&A...520A..43O,2012A&A...544A..32L,Levison2015,2017ASSL..445..197O}. Through these processes, embryos grow in mass and lead to the assembly of the terrestrial planets and the formation of giant planets’ cores.

As a planet grows, it alters the gas structure in the disc. Its gravitational potential excites perturbations at its Lindblad and corotation resonances. The net disturbance arising from the former takes the form of a spiral wake \citep{og2002,2018ApJ...859..118B} and yields a contribution to the torque named the Lindblad torque \citep{ww86,tanaka2002}. Likewise, the disturbance at corotation resonances gives rise to the corotation torque \citep{gt79,wlpi92}.  These resonant torques cause the embryo's radial migration \citep{tanaka2002,2014prpl.conf..667B}, which is generally directed inwards, and are responsible for the damping of its eccentricity and inclination \citep{arty93b,1994ApJ...423..581A,2004ApJ...602..388T,2011ApJ...737...37M}.

Another type of interaction with the disc, thermal forces, has been the focus of recent research. Thermal forces can arise from the diffusion of heat in the planet's vicinity \citep[][]{2014MNRAS.440..683L} or from the release of heat by the planet into the surrounding gas \citep[][]{2015Natur.520...63B}. Growing planets are indeed heated by the release of potential energy during differentiation processes and accretion impacts \citep{ElkinsTanton2012, Chao2021}. As a consequence of their surface's heating, planetary embryos have an effective luminosity that acts as a local heat source that perturbs the surrounding gas. The perturbation of density due to the release of energy, or radiative feedback, modifies the force exerted on the planet.
At low luminosity, thermal forces tend to damp the eccentricity and inclination of embryos much more vigorously than the resonant torques \citep{2019MNRAS.485.5035F,2017arXiv170401931E,2023MNRAS.tmp..658C}, while they also generally induce inwards migration \citep{2014MNRAS.440..683L,2017MNRAS.472.4204M}. When the luminosity to mass ratio of an embryo exceeds a critical value that depends on the local properties of the gas, thermal forces excite orbital eccentricities and inclinations \citep{2017arXiv170401931E, 2017A&A...606A.114C,
  2019MNRAS.485.5035F} and, for embryos with low eccentricity, give origin to a net positive torque which leads to outwards migration \citep{2015Natur.520...63B}. 

There has been a number of attempts to assess the impact of thermal forces on scenarios of planetary formation and migration by including expressions of the thermal torque \citep{2019MNRAS.486.5690G,Guilera2021,2020arXiv200400874B}.  Although some of them show the potential importance of radiative feedback on planetary growth, they consider only embryos on circular orbits and leave aside important processes, such as the evolution of eccentricity and inclination that have feedback on the accretion rate of pebbles \citep{2018A&A...615A.138L,2018A&A...615A.178O,2022MNRAS.509.5622V}. Besides, they consider the evolution of one embryo at a time, subjected to the disc forces and the accretion of pebbles or planetesimals.

Studies of pebble accreting embryos generally use an N-body module, together with prescriptions for pebble accretion and the interaction with the gas disc
\citep[e.g.][]{2014AJ....148..109K,2017A&A...607A..67M,2019A&A...623A..88B}. A common feature of this kind of study is that the gravitational force exerted by the gas onto the embryos is exclusively due to resonant interactions (that occur at Lindblad's and corotation resonances).

The main objective of this work is to assess the impact of thermal forces on the dynamics and growth of a set of low-mass planetary embryos.
We do this through N-body simulations in which the embryo-disc interactions are incorporated as additional forces, while we use the analytical expressions of \citet{2018A&A...615A.138L} and \citet{2018A&A...615A.178O} to evaluate the accretion rate (and thence the luminosity) of the embryos. We perform a large number of simulations spanning a $1$~Myr time frame by varying the random seed from one calculation to another. For each random seed, we perform two calculations: one in which the gas force acting on the embryos is only the resonant force exerted by the gas, and another one in which we also incorporate thermal forces. We make no pretence for realism or observational predictability. Far too little is known of the properties of the inner disc at the present time. A fine-grain knowledge of the value of the gradients of the disc's surface density and temperature as a function of radius is absolutely crucial to make quantitative predictions of planetary migration, and beyond the reach of present facilities for the disc's innermost astronomical units. We therefore resort to the customary ersatz of power law discs\footnote{Discs in which the surface density and temperature are power laws of the distance to the star.}, in which we insert an \emph{ad hoc} migration trap. While probably unrealistic in many aspects, our setup provides a minimal environment to study how the introduction of thermal forces can affect the dynamics and the growth of a system of embryos, by comparing the outcome of simulations that include these forces to the outcome of simulations that do not.

Our work is organised as follows: in Section~\ref{sec:methods} we describe the N-body code that we use, and lay out our disc model. In Section~\ref{sec:evol-an-isol} we perform a preliminary study of the orbital evolution and mass growth of an isolated embryo, in order to get a sense of the impact of thermal forces. We will see, in particular, that these forces act on a much shorter timescale than the forces arising from the resonant interaction with the disc. In Section~\ref{sec:evol-swarms-embry}, we turn to the study of the evolution of a set of embryos. We consider two different kinds of initial distributions: one strongly packed and another one with more distant embryos. In Section~\ref{sec:discussion} we discuss our main findings, and draw a list of the main shortcomings and caveats of our study, before concluding in Section~\ref{sec:conclusions}. 

\section{Methods}
\label{sec:methods}
We summarise hereafter the main features of our implementation, for the N-body part and for the underlying disc model. The full detail is given in Appendix~\ref{sec:formulaciones}.
\subsection{N-body solver}
\label{sec:n-body-solver}
We use the N-body code \texttt{REBOUND} \citep{2012A&A...537A.128R} with the integrator \texttt{IAS15} \citep{IAS15}. Planet-disc interactions are included by modifying the acceleration exerted on each embryo using analytical prescriptions. This means that each embryo is subjected to both embryo-disc interactions and interactions with other embryos:
\begin{equation}
  \label{eq:1}
\frac{d^{2} \mathbf{r}_{i}}{dt^{2}} =  -GM \frac{\mathbf{r}_{i} - \mathbf{r}_{\ast}}{ \lvert \mathbf{r}_{i} - \mathbf{r}_{\ast}\rvert^{3}} - G \sum\limits_{j \neq i}^{N} m_{j} \frac{\mathbf{r}_{i} - \mathbf{r}_{j}}{ \lvert \mathbf{r}_{i} - \mathbf{r}_{j}\rvert^{3}} + \frac{1}{m_{i}} \left( \mathbf{F}^{[R]}_i + \mathbf{F}^{[T]}_i\right)
\end{equation}
where $\mathbf{r}_i$ is the position vector of embryo $i$ in an inertial frame, $m_i$ its mass, $G$ the gravitational constant, $M$ the mass of the central object and $\mathbf{r_\ast}$ its position, $N$ the number of embryos, $\mathbf{F}^{[R]}_i$ is the force on embryo $i$ arising from resonant interaction with the disc and $\mathbf{F}^{[T]}_i$ the thermal force on embryo $i$.
The expressions for these forces are given respectively in Appendices~\ref{sec:resonant} and~\ref{sec:thermal}.

Additionally, our code:  
\begin{enumerate}
\item Allows (perfect) mergers between embryos while preserving the mass, momentum and volume of the colliding planetary embryos.
\item Includes formulae that describe pebble accretion. These expressions take into account the orbital parameters of each embryo following the prescriptions of
  \citet{2018A&A...615A.138L} and \citet{2018A&A...615A.178O}, which are described in Appendix~\ref{sec:pebble-accretion}. We furthermore switch off the accretion of pebbles onto a given embryo once its mass exceeds the pebble isolation mass given by \citet{Ataiee2018}.
\item Evaluates the luminosities $L_i$ of the embryos arising from the accretion rates $\dot m_i$. We use the expression:
  \begin{equation}
    \label{eq:2}
    L_i=\frac{Gm_i\dot m_i}{R_i},
  \end{equation}
  where $R_i$ is the physical radius of embryo $i$, determined assuming a density $\rho_\bullet=3$~g.cm$^{-3}$ for all our solid bodies. The luminosities, in turn, are used to evaluate the thermal forces.
\end{enumerate}
\subsection{Disc model}
\label{sec:disc-model}
The  embryos evolve in a gaseous disc in which the surface density $\Sigma$ and temperature $T$ are power laws of the distance $r$ to the central object (a solar mass star):
\begin{equation}
  \label{eq:3}
\frac{\Sigma(r)}{\mathrm{g.cm^{-2}}}=200\left(\frac{r}{4\;\mathrm{au}}\right)^{-1/2}
\end{equation}
and
\begin{equation}
  \label{eq:4}
\frac{T(r)}{\mathrm{K}}=126\left(\frac{r}{4\;\mathrm{au}}\right)^{-1}.
\end{equation}
The law of temperature implies that the snow line, reached for $T\simeq 170$~K, is located at $r\simeq 3$~au.
The laws of surface density and temperature are constant in time: we do not track the evolution of the disc. However, because turbulence has an impact on the accretion of pebbles \citep{2018A&A...615A.178O}, we assume that our disc has an $\alpha$-viscosity\footnote{Since, for our purpose, we are only interested in the effects of the vertical turbulent diffusivity, we use the notation $\alpha_z$, following \citet{2018A&A...615A.178O}.} \citep{ss73}, with $\alpha_z=10^{-4}$. A key parameter in the evaluation of thermal forces is the thermal diffusivity of the disc, given by \citep[][and refs therein]{2017MNRAS.471.4917J}:
  \begin{equation}
    \label{eq:14}
    \chi = \frac{16 \left( \gamma - 1 \right) \sigma T^{3} \mu }{3 \rho_{0}^{2} \mathcal{R} \kappa},
  \end{equation}
  where $\sigma$ is the Stefan–Boltzmann constant, $\rho_0$ the midplane density, $\mathcal{R}$ the constant of ideal gases, $\mu$ the mean molar mass and $\kappa$ the opacity of the gas. We calculate the latter using the prescription of \citet{1994ApJ...427..987B}.

  In addition, we call $\gamma$ the ratio of specific heats of the gas, $c_s$ the adiabatic sound speed, given by:
  \begin{equation}
    \label{eq:41}
    c_s=\sqrt\frac{\gamma\mathcal{R}T}{\mu},
  \end{equation}
  and $H$ the vertical pressure length scale:
  \begin{equation}
    \label{eq:42}
    H=\frac{c_s}{\gamma^{1/2}\Omega}.
  \end{equation}
  The aspect ratio $h$ of the disc is the ratio of the pressure length scale to the local distance to the star:
  \begin{equation}
    \label{eq:43}
    h=\frac{H}{r}.
  \end{equation}
  Our temperature law implies that the aspect ratio of the disc is constant and has value $h\equiv 0.045$.
  The density of the gas at the disc's midplane is:
  \begin{equation}
    \label{eq:44}
    \rho_0(r)=\frac{\Sigma(r)}{\sqrt{2\pi}H},
  \end{equation}
  and the density at an arbitrary location $(r,z)$ is given by:
  \begin{equation}
    \label{eq:45}
    \rho(r,z)=\rho_0(r)\exp(-z^2/2H^2).
  \end{equation}
 The disc is traversed by inflowing pebbles, which have a radial mass flux $\dot M_\mathrm{peb}=10^{-4}(Z/Z_\odot)\;M_\odot.\mathrm{yr}^{-1}$ at large distance (i.e. beyond the outermost embryo) and a unique Stokes number $\tau_s=10^{-2}$. When we vary the disc's metallicity, in addition to updating $\dot M_\mathrm{peb}$, we also multiply $\kappa$ by $Z/Z_\odot$. We consider five values of the metallicity: $Z=0.50 Z_\odot$, $0.75 Z_\odot$, $1.0 Z_\odot$, $1.25Z_\odot$ and $Z=1.50 Z_\odot$.

\subsection{Special treatment of the snow line}
\label{sec:spec-treatm-snowl}
We do not use the full formulae for the torques exerted on the embryos, which would take into account the different components of the corotation torque \citep[][and references therein]{2017MNRAS.471.4917J} and their degrees of saturation. Rather, we use the expressions of \citet{tanaka2002} and \citet{2004ApJ...602..388T} for the resonant torques. These expressions are those of the torque exerted on a planet embedded in a globally isothermal disc (i.e., a disc with a uniform temperature) in which the surface density is a power law of radius. While they give a reasonable order of magnitude of the inward migration timescales of the embryos in the smooth parts of the disc, applying these torques to forming embryos is a well-known recipe to rush them toward the star on a short timescale \citep{Ida:2008p396}. The power law approximation to the disc profiles is an oversimplification that does not capture the existence of migration traps that exist where embryos are subjected to strong, positive corotation torques, such as those due to an entropy gradient \citep{2010ApJ...715L..68L, 2015MNRAS.452.1717L} or to a vortensity gradient \citep{trap06}. The snow line, in particular, can act as a migration trap when a small-scale thermal structure appears as a result of a transition in the fragmentation velocity of pebbles \citep{2021A&A...650A.185M}, or because small-scale variations of the temperature and surface density can lead to large vortensity gradients \citep{2011CeMDA.111..131M}. For this reason, in addition to the simple prescription for the torques mentioned above, we manually add an \textit{ad hoc} migration trap at the snow line by incorporating an extra, positive contribution to the torque wherever the disc's temperature is close to $170$~K, the approximate temperature of the sublimation of water ice \citep[see also ][for an alternative implementation of an \textit{ad hoc} migration trap]{2014AJ....148..109K}. The procedure is described in Appendix~\ref{sec:limit-low-mach}.

\section{Evolution of an isolated planetary embryo}
\label{sec:evol-an-isol}
In this section we present preliminary calculations performed for one embryo only, in order to highlight how the introduction of thermal forces affects its orbital evolution and mass growth with respect to the case in which the embryo is subjected only to the classical resonant forces (arising from Lindblad's and corotation resonances).

\begin{figure}
 \includegraphics[width=\columnwidth]{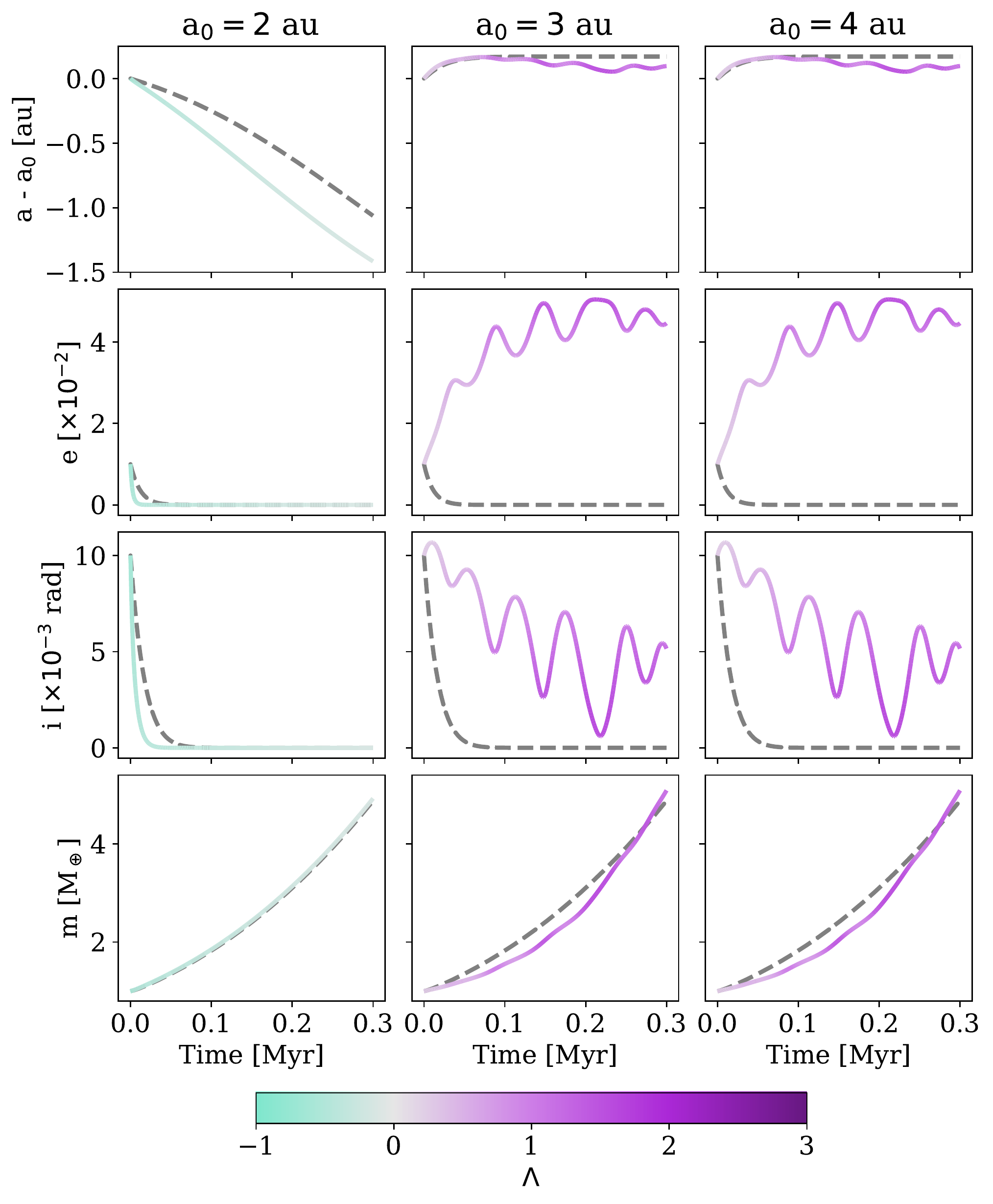}
 \caption{Orbital evolution of a planetary embryo with initial mass $1.0 M_{\oplus}$ starting at different semi-major axes $a_0$. The initial eccentricity $e_0$ and inclination $i_0$ are both equal to $0.01$. The evolution of the embryo under resonant interactions only is shown with a dashed line, while the evolution with the contribution of thermal interactions is shown with a solid line and a colour that reflects the value of $\Lambda$.}
 \label{fig:EvolucionOrbital}
\end{figure}

In a first set of experiments, we evolved a planetary embryo of mass $m=1.0 M_{\oplus}$ for $300$~kyr starting from different semi-major axes ($2$~au, $3$~au and $4$~au) with initial eccentricity and inclination $e_{0} = i_{0}= 0.01$. The results of these experiments are presented in Figure \ref{fig:EvolucionOrbital}. As mentioned in Section~\ref{sec:introduction}, thermal forces change sign for a critical value of the luminosity-to-mass ratio. This dependence can be described by a dimensionless parameter $\Lambda$, the expression of which is given in detail in the Appendix~\ref{sec:thermal-force-shear}. When $\Lambda < 0$, the embryo is said to be ``cold'' and the effects of thermal diffusion on the perturbed flow dominate over those of heat release, whereas when $\Lambda > 0$, the effects of heat release are dominant and the embryo is qualified of ``hot''. The case $\Lambda=0$ corresponds to the critical luminosity-to-mass ratio, at which thermal effects cancel out.

We find that embryos interior to the snow line (left column of Fig.~\ref{fig:EvolucionOrbital}) remain cold. As a consequence, they migrate inward faster than they would if subjected to resonant forces only (as the ``cold finger effect'' of \citet{2014MNRAS.440..683L} is the dominant thermal effect on these objects), while their eccentricity and inclination are also damped on a shorter timescale \citep{2019MNRAS.485.5035F}. Despite this different orbital evolution with respect to the case where thermal forces are discarded, the evolution of their mass is virtually indistinguishable.

Embryos starting at the snow line or beyond (shown respectively in the middle and right columns of Fig.~\ref{fig:EvolucionOrbital}) eventually become hot. Under these circumstances, their eccentricity is excited and saturates toward a value comparable to the aspect ratio of the disc \citep{2022MNRAS.509.5622V}. While their inclination slightly increases over the first $\sim 10$~kyr, it subsequently reverses its trend and decays in an oscillatory manner, so that the end state is that of an embryo with a sizeable eccentricity and a moderate inclination. These results are compatible with the long-term runs of \citet{2017arXiv170401931E}, which span shorter time scale as they were obtained from computationally expensive hydrodynamical numerical simulations.
Note that the stalled migration of the embryo at the snow line simply occurs because of the \textit{ad hoc} trap that we set there.
The embryo released at $4$~au migrates inwards and is eventually trapped at the snow line. It initially migrates faster if subjected only to resonant interactions, a result compatible with the findings of \citet{2015Natur.520...63B}.
\begin{figure}
  \includegraphics[width=\columnwidth]{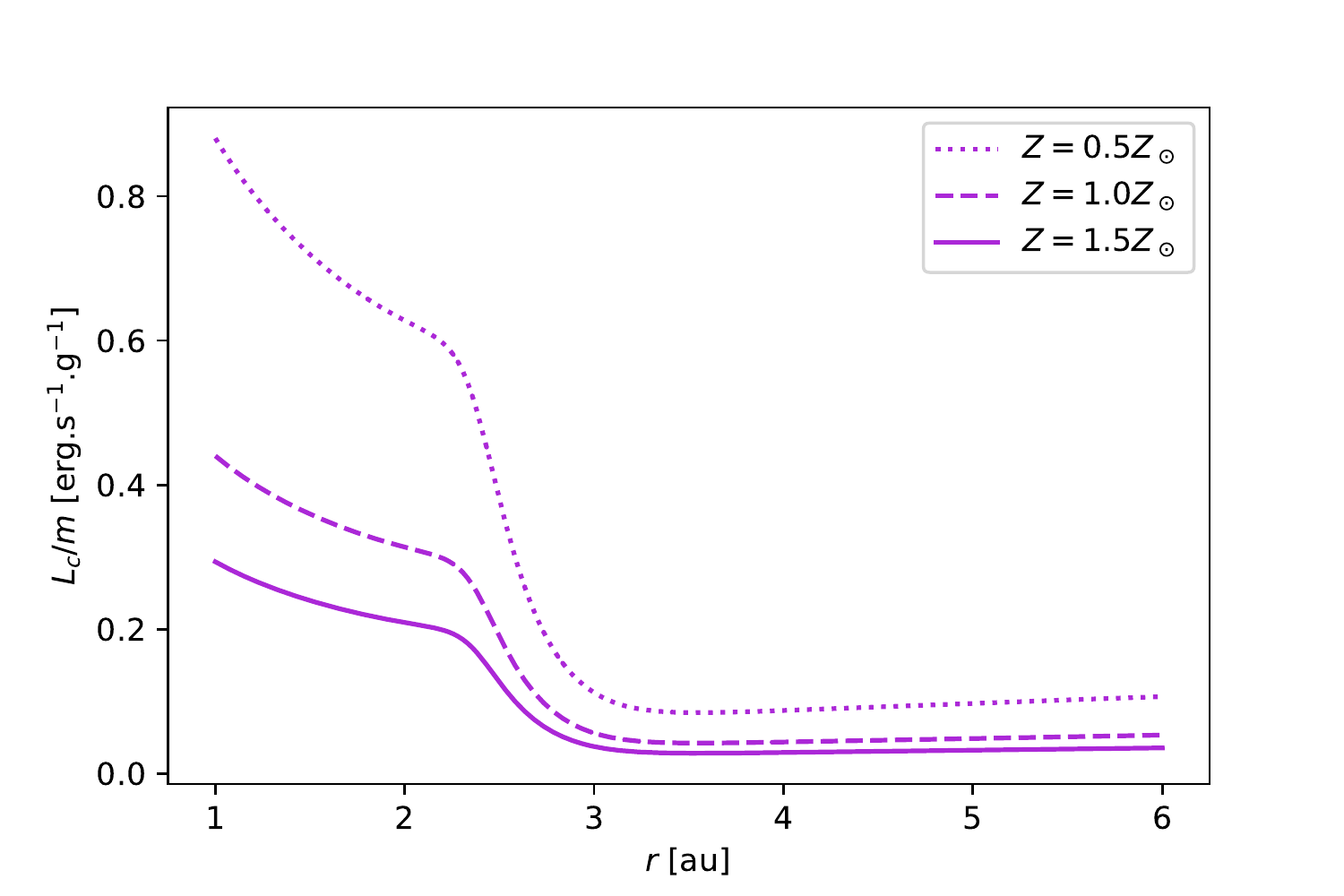}
  \caption{Critical luminosity to mass ratio as a function of radius, for different values of the metallicity considered in this work. An embryo that has a luminosity to mass ratio above the critical value undergoes an excitation of eccentricity and inclination (it is then qualified of ``hot''), whereas these quantities are damped if its luminosity to mass ratio is below the critical value (it is then qualified of ``cold'').}
  \label{fig:lcoverm}
\end{figure}
\begin{figure*}
 \includegraphics[width=\textwidth]{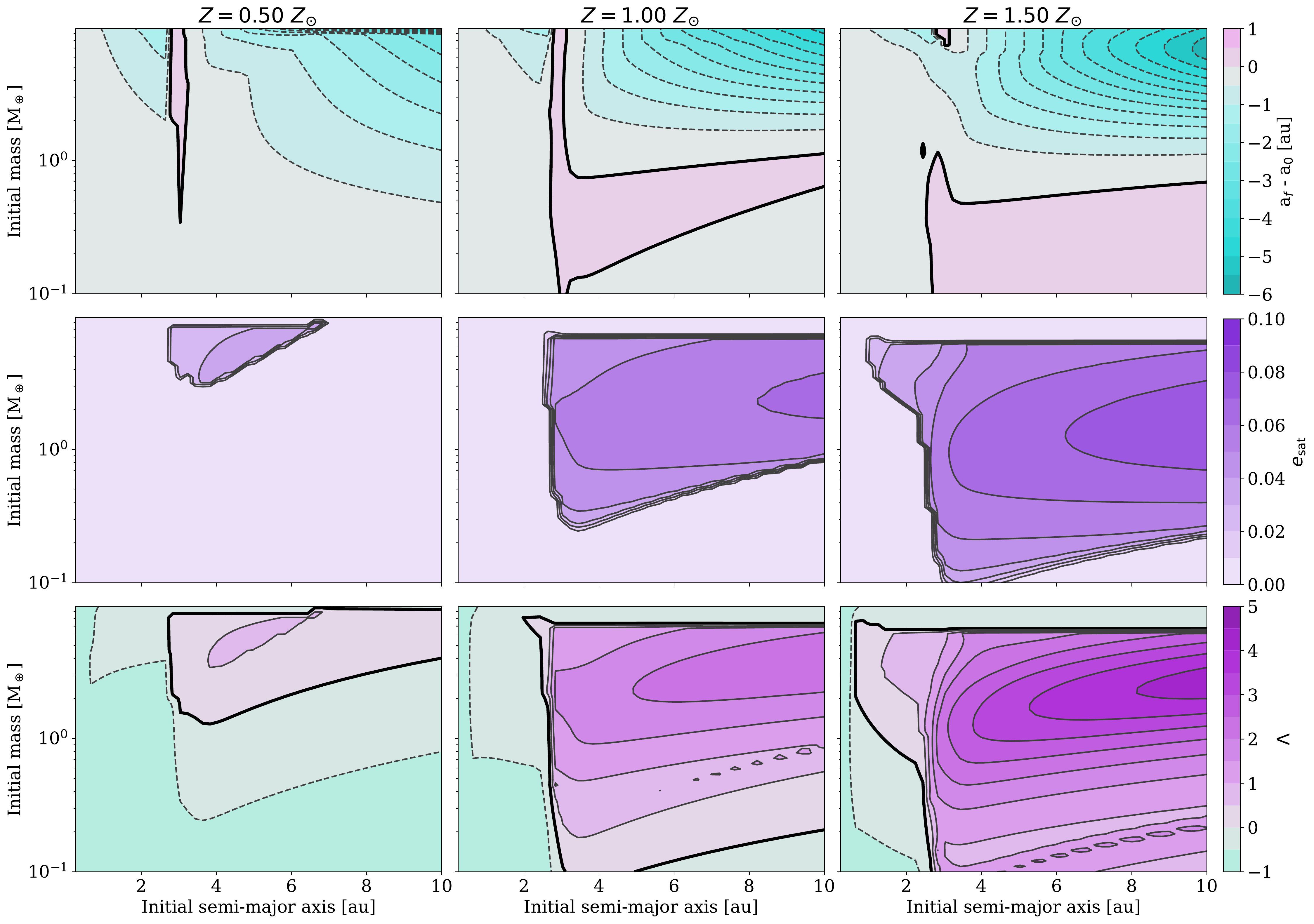}
 \caption{Orbital evolution over $50$~kyr and final thermal status of single embryos, as a function of their initial semi-major axis and their initial mass, for the five different values of metallicities mentioned in Section~\ref{sec:disc-model}. From top to bottom: (i) net variation of the semi-major axis $a_f-a_0$; (ii) eccentricity reached at $50$~kyr (this timescale being much larger than that of the eccentricity's variation, the value plotted can be regarded as the asymptotic value) and (iii) final thermal status of the embryo, as given by the dimensionless factor $\Lambda$. In this last plot, the turquoise region corresponds to embryos which are dominated by the cold force, whereas the violet region corresponds to  embryos dominated by the heating force.}
 \label{fig:Mapa-Fiducial}
\end{figure*}

The ``dichotomy'' between the embryos inside and outside the snow line can be understood by examining the critical luminosity to revert thermal forces, per unit planetary mass (Fig.~\ref{fig:lcoverm}). The critical luminosity is much higher inside the snow line, explaining why embryos tend to remain cold in this region of the disc. Also, the critical luminosity decreases as the metallicity increases. The values of Fig.~\ref{fig:lcoverm} should be compared to the luminosity-to-mass ratio of an accreting, low-mass embryo, obtained using Eq.~\eqref{eq:2}:
\begin{equation}
  \label{eq:46}
  \frac Lm=0.16\left(\frac{m}{M_\oplus}\right)^{2/3}\left(\frac{\rho_\bullet}{3\;\mathrm{g.cm}^{-3}}\right)^{1/3}\left(\frac{\tau}{10^5\;\mathrm{yr}}\right)^{-1}\;\mathrm{erg.s}^{-1}.\mathrm{g}^{-1},
\end{equation}
where $\tau=m/\dot m$ is the mass doubling time. We see that an Earth-mass embryo with a mass doubling time of $10^5$~yr has a supercritical luminosity outside of the snowline, for all the metallicity values considered, and conversely that it has always a subcritical luminosity inside the snowline, regardless of the metallicity.

Finally, we studied the evolution of single embryos subjected to thermal forces over $50$~kyr as a function of their initial semi-major axis and initial mass (ranging from $0.1 M_{\oplus}$ to the pebble isolation mass value given by \citet{Ataiee2018}, which is approximately $10$~$M_\oplus$ for the parameters that we use in our simulations, regardless of the distance to the star). Their orbital and thermal behaviour is shown in the maps of Fig.~\ref{fig:Mapa-Fiducial}.
They show that the domain of outward migration is more extended at larger metallicity (on the one hand, the accretion rate of the embryos, hence their luminosity, increases with the metallicity while on the other hand the thermal diffusivity of the gas decreases, lowering the threshold luminosity for the embryos to become ``hot'' and migrate outwards, as can be seen in Fig.~\ref{fig:lcoverm}). Likewise, the domain with a  finite asymptotic eccentricity  grows as the metallicity increases. Finally, the dimensionless luminosity increases  with the metallicity, for any mass and semi-major axis. The plots of Fig.~\ref{fig:Mapa-Fiducial} also show that the maximum eccentricity is attained at lower mass for higher metallicity, and that the upper boundary of the region of outward migration, similarly, shifts toward lower masses. The reason for this behaviour arises from the cut-off of thermal effects (presented in Appendix~\ref{sec:thermal}), which occurs for planetary masses beyond a critical mass that scales with the disc's thermal diffusivity \citep{2020MNRAS.495.2063V}. All our discs having the same temperature profile, the thermal diffusivity scales as the inverse of the opacity (or metallicity) and so does the critical mass.
\begin{figure}
  \centering
  \includegraphics[width=.85\columnwidth]{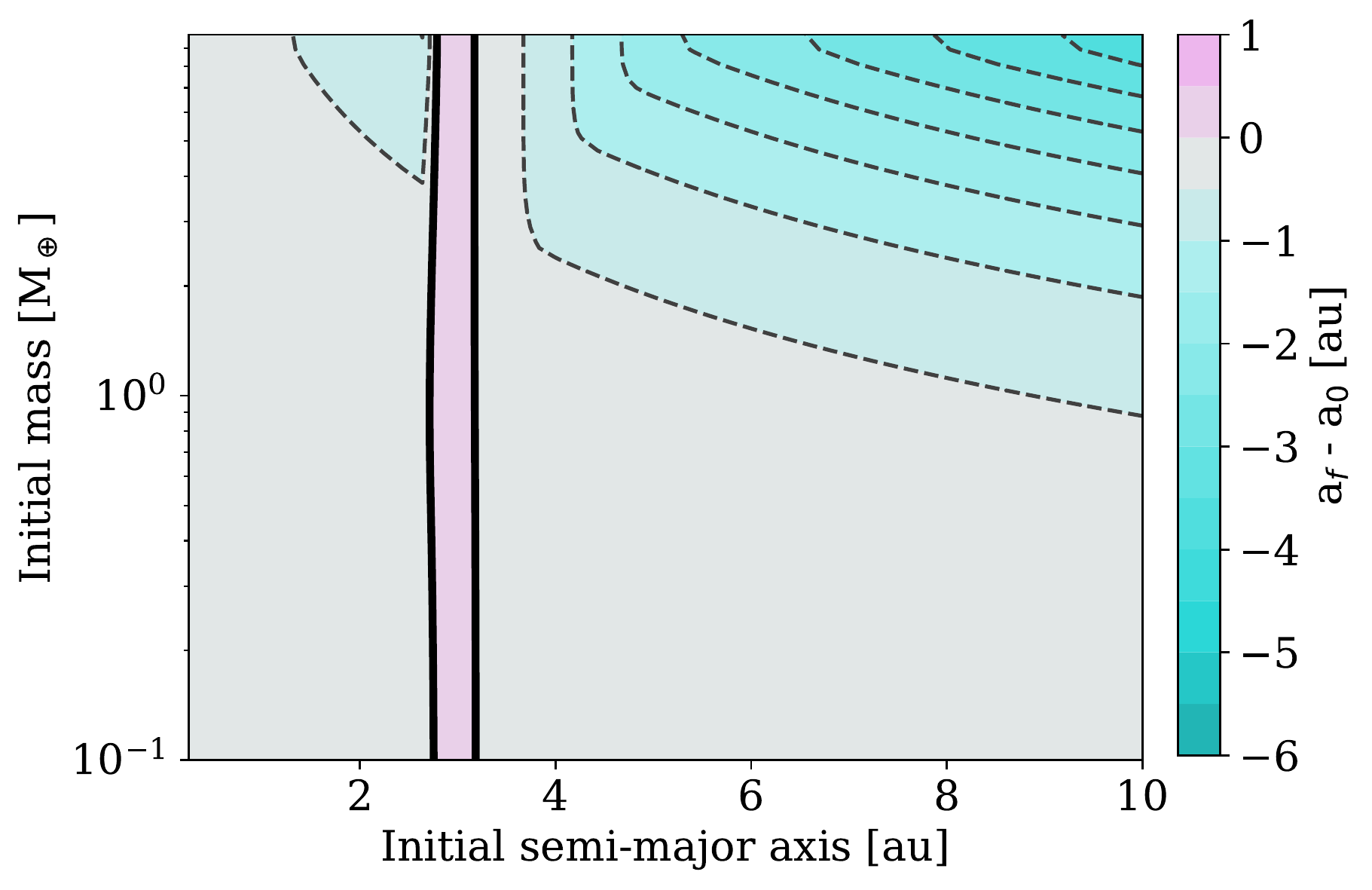}
  \caption{Orbital evolution in the purely resonant case. This map has been obtained for the metallicity $Z=Z_\odot$, but the small mass increase over the short time span of this run implies that this map is almost identical for all the values of the metallicity considered here.\label{fig:mapas}}
\end{figure}
This behaviour is to be contrasted with that of the purely resonant case, shown in Fig.~\ref{fig:mapas}.
Except for a narrow region around the \textit{ad hoc} migration trap, migration is only inward. The final eccentricity is not shown for this case, as it would be zero everywhere.

\section{Evolution of sets of embryos}
\label{sec:evol-swarms-embry}
\subsection{Two different kinds of setups}
\label{sec:two-different-kinds}
For each value of the metallicity mentioned in Section~\ref{sec:disc-model}, we consider $30$ sets of embryos that at $t=0$ all have a mass of $0.1\,M_\oplus$. This is an arbitrary but widespread assumption, which is discussed and relaxed by \citet{2022A&A...666A..90V}, who consider the formation of embryos in a more self-consistent fashion. The eccentricities and inclinations of the embryos have random values given by Rayleigh distributions with dispersions $\sigma_{e} = 10^{-2}$ and $\sigma_{i} = 5\cdot 10^{-3}$ respectively \citep{Lissauer1993}. For each set we perform two simulations that cover each a $1$~Myr time span: (i) one in which the embryos are evolved exactly as described in Section~\ref{sec:methods} and (ii) another one with identical prescriptions except for the fact that we set $\mathbf{F}^{[T]}_i$ to zero in Eq.~\eqref{eq:1}. The first case corresponds, therefore, to the inclusion of both resonant and thermal forces, whereas the second case corresponds to the inclusion of resonant forces only.

We consider two different distributions of semi-major axis at $t=0$:
\begin{enumerate}
\item One in which $N=40$ embryos are equally spaced from $a=3$~au to $a=6$~au.
\item One in which $N=16$ embryos are logarithmically spaced from $a=3$~au to $a=6$~au.
\end{enumerate}
The first distribution corresponds to a dynamically packed set of embryos, with separations that bracket the stability criterion  of \citet{gladman93} (this criterion is satisfied near $a=3$~au and violated near $a=6$~au). We will refer to this kind of set as dynamically active \citep{2008ApJ...686..603J}. The second distribution corresponds to embryos that are separated by a fixed number $\Delta=8$ of mutual Hill radii \citep{2002ApJ...581..666K}, where the mutual Hill radius is given by \citep{1986Icar...66..536P}:
\begin{equation}
  \label{eq:9}
  R_H=\bar a \left(\frac{2m}{3M}\right)^{1/3},
\end{equation}
where $\bar a$ is the mean of the semi-major axes of the embryos.
We call this kind of distribution dynamically quiet. This distinction is somewhat arbitrary, as both sets eventually undergo dynamical relaxation. It is simply meant to refer concisely to either set.

\subsection{Run results}
\label{sec:dynam-active-embry}
We show in Figs.~\ref{fig:Resumen-Metalicidad} and~\ref{fig:Resumen-Metal2} the distribution of semi-major axis, eccentricities and inclinations of the embryos of all our simulations after $1$~Myr, for the extreme and middle values of the metallicity considered here. The results for dynamically active sets are shown in the left plots and those for dynamically quiet sets in the right plots. This figure shows that:
\begin{figure*}
   \includegraphics[width=0.85\textwidth]{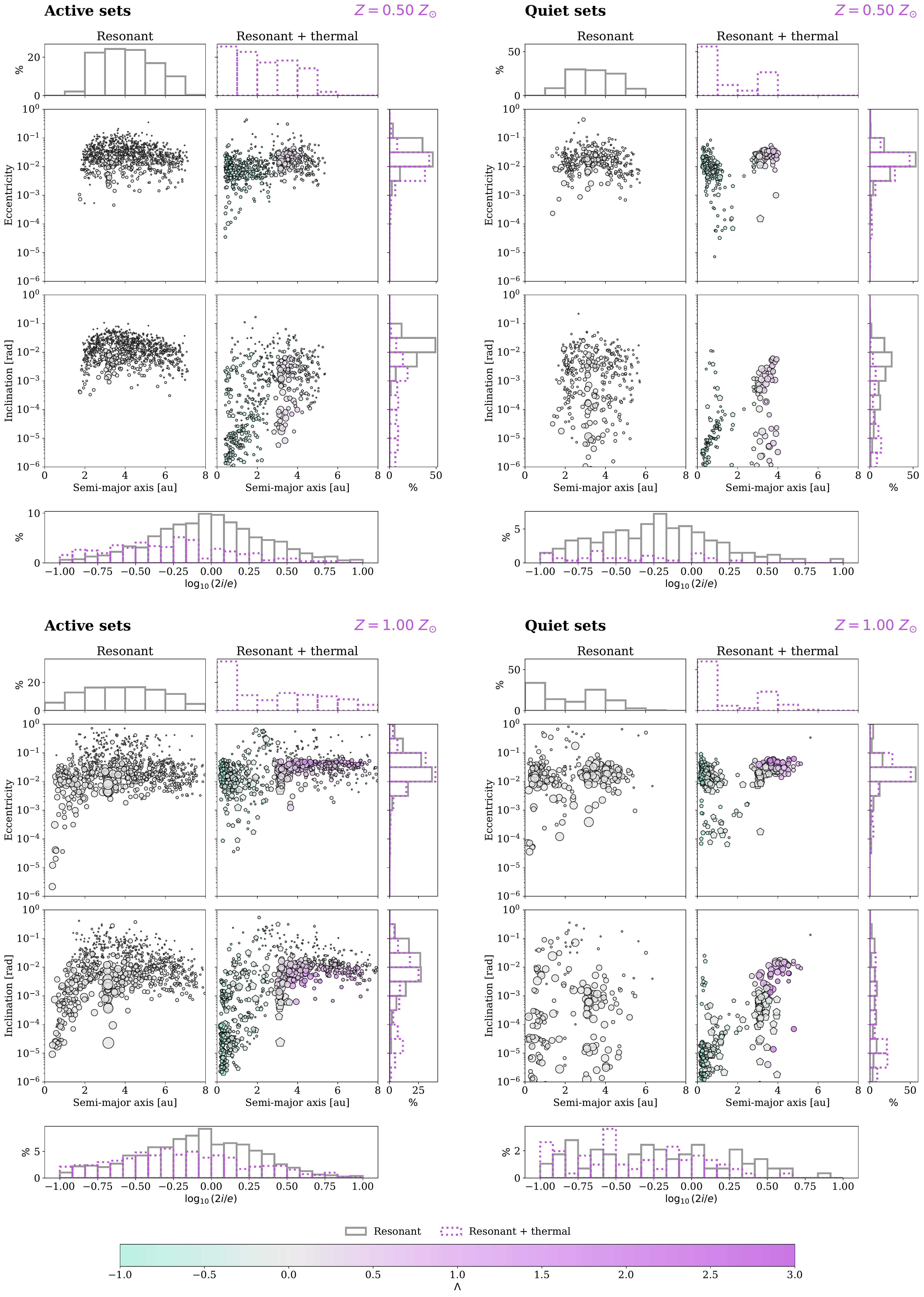}
 \caption{Distribution of eccentricities and inclinations as a function of semi-major axis, for the active set (left) and for the quiet set (right), at $t=1$~Myr, for metallicities $Z=0.5Z_\odot$ (top) and  $Z=Z_\odot$ (bottom). For each plot, the left column shows the case with resonant forces only, while the right column shows the case with both resonant and thermal forces. In that case the colour of each symbol corresponds to the dimensionless luminosity of the embryo (using the colour bar at the bottom of the figure). The projected histograms show the distribution of semi-major axis (top row) and eccentricity and inclination (on the right side). The size of the symbols scales linearly with the mass. Under each plot, a histogram shows the distribution of $\log_{10}\left( 2i/e \right)$. The radius of a symbol scales with the mass of the corresponding embryo. The solid grey lines show the data for runs with resonant forces only, whereas the dashed purple lines show the data for runs with resonant and thermal forces.
    \label{fig:Resumen-Metalicidad}}
\end{figure*}
\begin{figure*}
  \centering
  \includegraphics[width=0.85\textwidth]{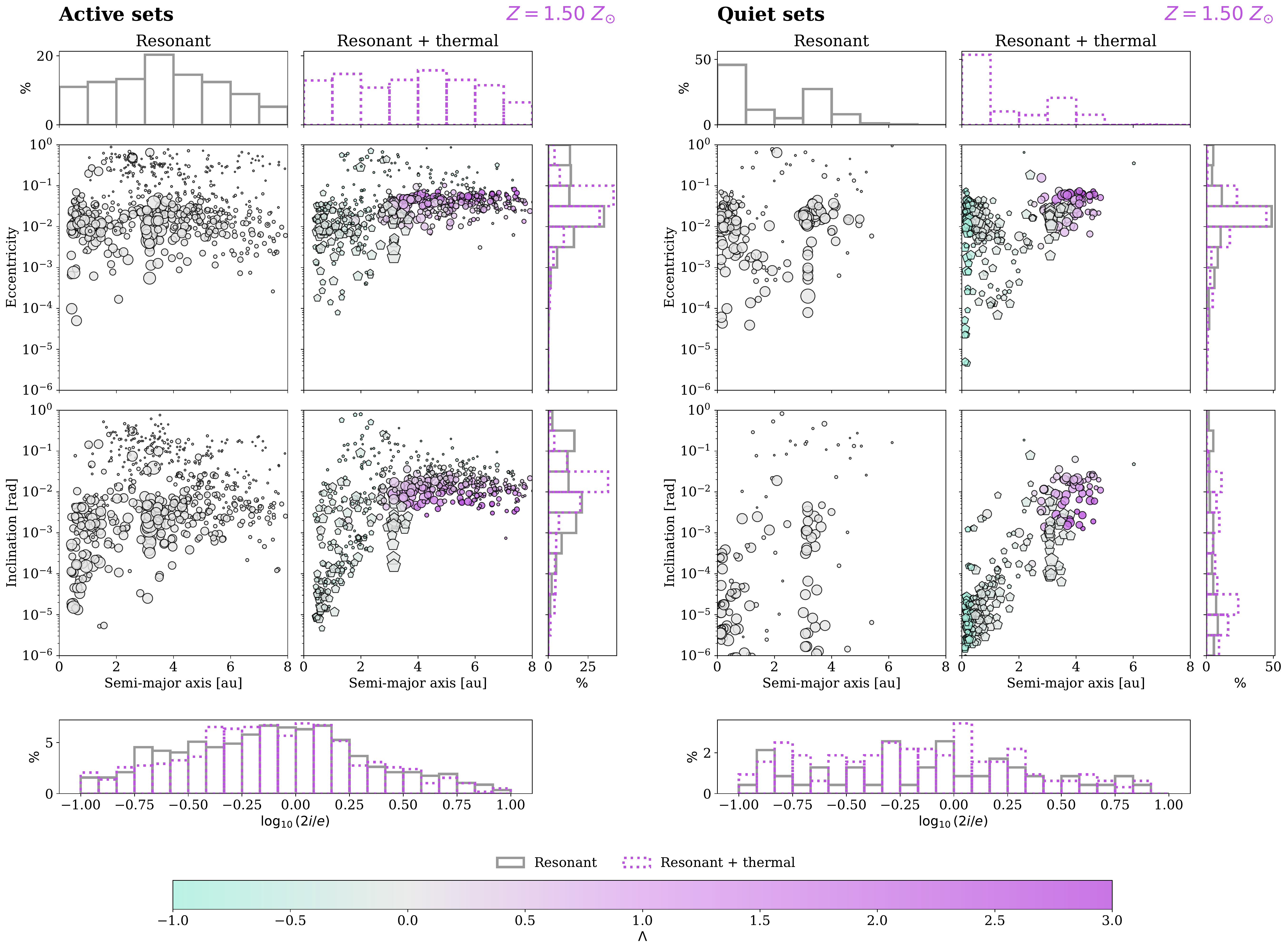}
  \caption{Same as Fig.~\ref{fig:Resumen-Metalicidad} for metallicity $Z=1.5Z_\odot$.}
  \label{fig:Resumen-Metal2}
\end{figure*}
 \begin{figure*}
   \includegraphics[width=\textwidth]{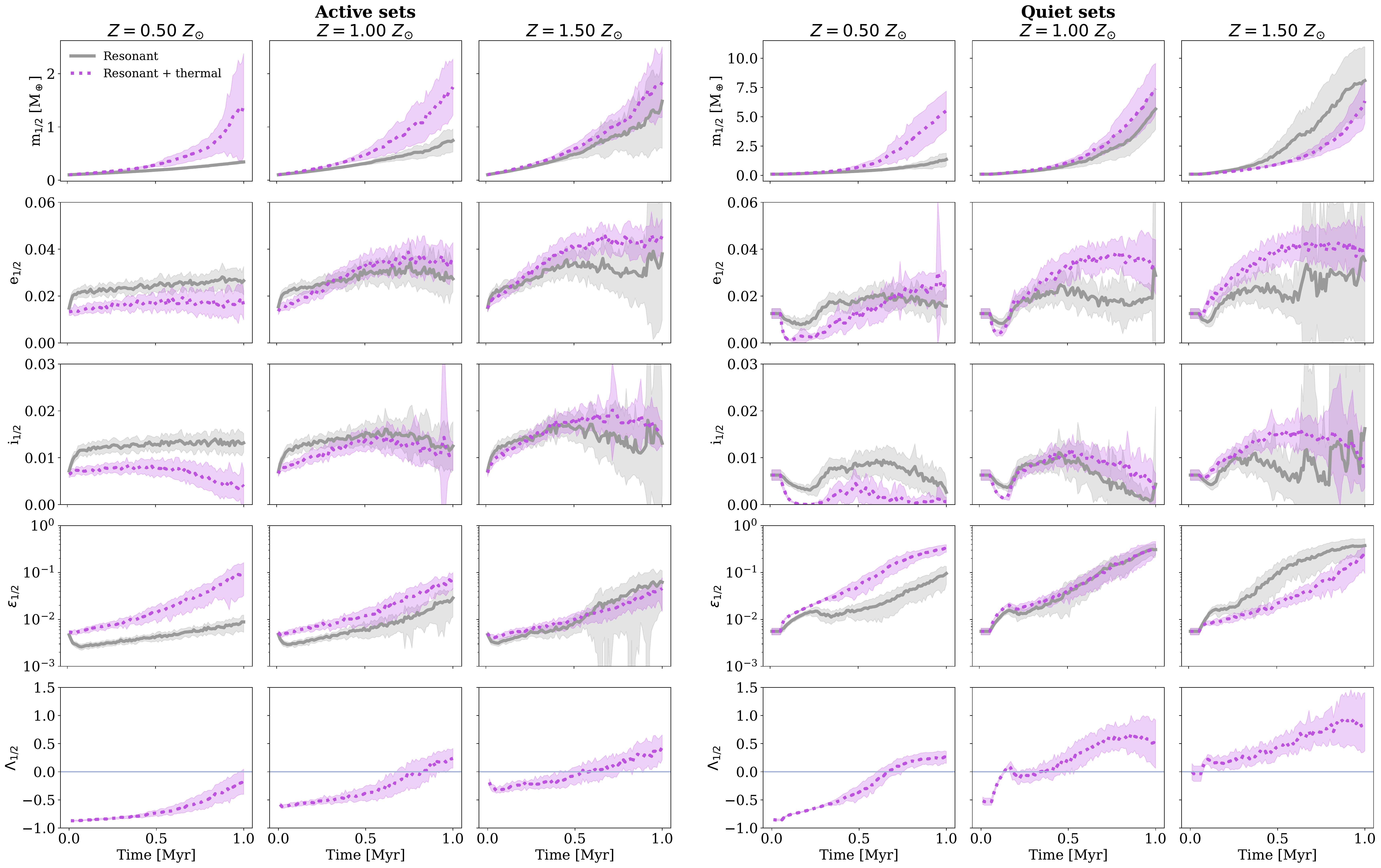}
   \caption{Ensemble average of the median eccentricities (top row), median inclinations (second row), median mass (third row), median filtering efficiency (fourth row), and dimensionless luminosity (bottom row), as a function of time, for the extreme and middle  values of the metallicity considered in our runs. The left panel shows the case of the dynamically active sets, and the right panel the case of the dynamically quiet sets. The dotted line (purple in the electronic version) shows the data for the runs with both resonant and thermal forces, and the solid line (grey in the electronic version) shows the data for the runs with the resonant forces only. In the top four rows, the shaded areas depict the dispersion of the values (the height of these areas is equal to the [ensemble average of the] root mean square deviation of the variable under consideration).}
   \label{fig:synoptic}
 \end{figure*}
\begin{enumerate}
\item For each value of the metallicity, there is a marked difference between the outcome of the runs with resonant forces only (left column of a plot) and the runs with both resonant and thermal forces (right column of a plot).
\item The embryos outside the snow line tend to be hot (i.e. have a supercritical luminosity) when thermal forces are included, and have a more narrow dispersion of eccentricities than their counterpart in runs with resonant forces only.
\item The histograms displaying $\log_{10}\left( 2i/e \right)$ show a deviation from equipartition between $e$ and $i$, particularly at lower metallicity, when embryos are subjected to thermal forces: $i$ is significantly  smaller than $e/2$ for these systems. The histograms of the sets subjected to resonant forces only are roughly centred around zero, as expected from equipartition \citep{2000Icar..143...28S}.
\item The embryos interior to the snow line, when thermal forces are taken into account, tend to cluster much closer to the star than when they are not. This is to be expected. As we saw already in Section~\ref{sec:evol-an-isol}, embryos within the snow line tend to be ``cold'' as the critical luminosity above which eccentricity and inclination are excited is large. As a consequence, they tend to have lower eccentricities and inclination than beyond the snow line. For a cold planet on a nearly circular orbit, the thermal force reduces to the ``cold finger effect'' of \citet{2014MNRAS.440..683L}, which, if the disc is sub-Keplerian (as is our power law disc everywhere), induces an inward migration which is faster, by a factor of several, than type~I migration \citep{2017MNRAS.472.4204M}.
\end{enumerate}
We shall now restrict our discussion and analysis to the outer disc for the rest of this paper, i.e. to the region exterior to $3$~au. The region interior to the snow line has a markedly different thermal diffusivity, and as a consequence, the behaviour of the embryos and their orbital statistics are significantly different between the inner embryos and the outer ones: it would make no sense to discuss both populations as a whole. Besides, we do not take into account in our simulations a possible change of size (and of Stokes number) of the dust when it crosses the snow line \citep{2015Icar..258..418M}, nor do we consider the possibility of dust capture at a pressure maximum in the vicinity of the snow line \citep{2021A&A...652A..35C}. For these reasons, in spite of all the simplifying assumptions of our model, we consider the outcome in the outer disc to be somehow more realistic.
\begin{figure}
\includegraphics[width=\columnwidth]{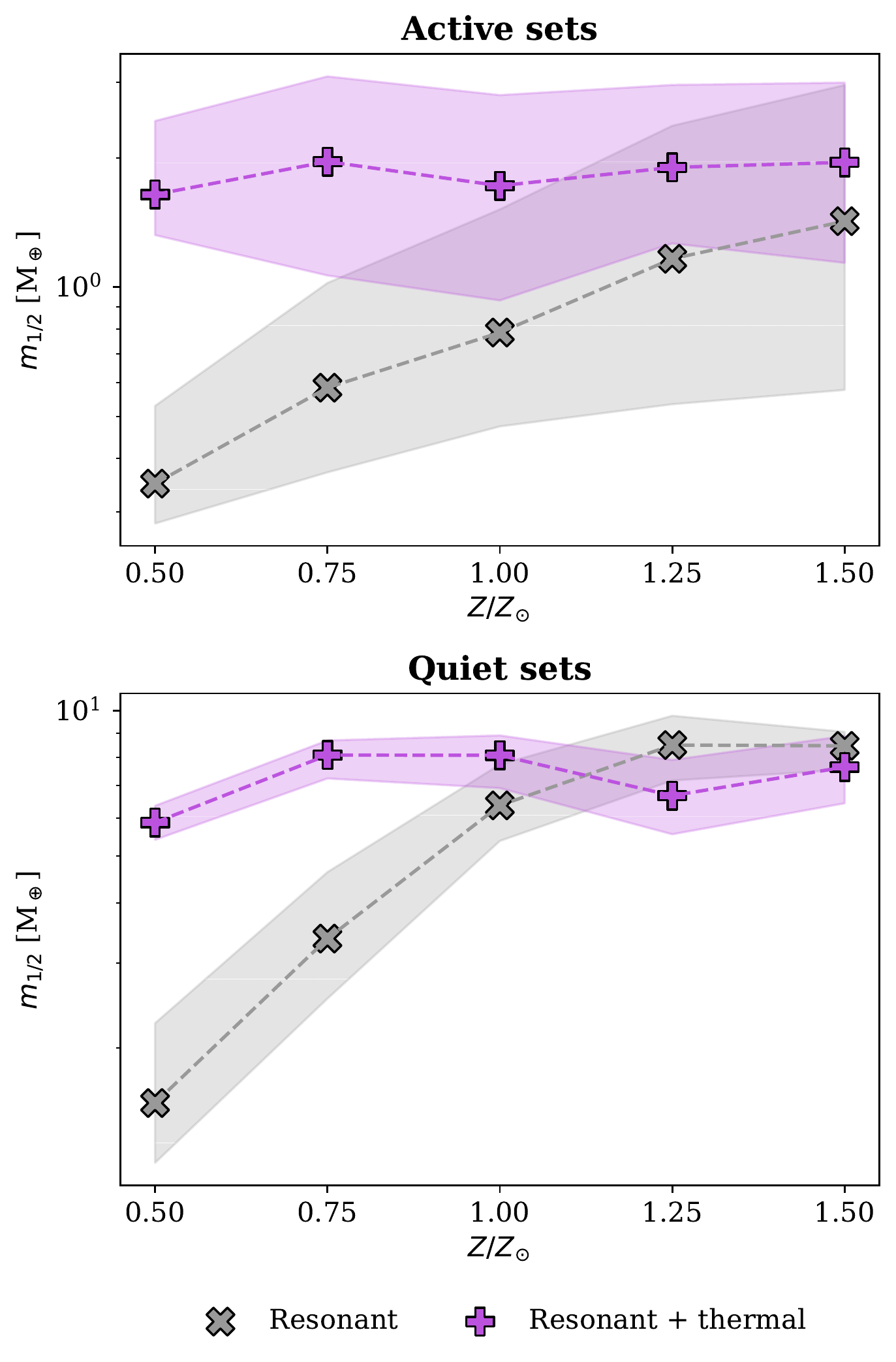}
  \caption{\label{fig:Median-Metalicidad}Ensemble averages of the median mass of embryos with semi-major axis $a>3$~au at the end of our runs as a function of metallicity, for the cases with all forces and the cases with the resonant force only. The top plot shows the results obtained with the active set, and the bottom plot shows the results obtained with the quiet set. The shaded areas are limited by the ensemble averages of the 32 and 68 percentiles, and give a sense of the typical mass dispersion obtained in each case.}
\end{figure}
In Fig.~\ref{fig:synoptic}, we show the ensemble averages of the median masses, eccentricities, inclinations, accretion efficiencies and dimensionless luminosities as a function of time, for all embryos that have $a>3$~au at a given time. The accretion (or filtering) efficiency $\varepsilon$ is defined as the mass fraction of the inflowing pebbles that is accreted by an embryo (see Appendix~\ref{sec:pebble-accretion}).  The median mass as a function of time shows a behaviour that hardly depends on the metallicity when thermal forces are included (although the behaviour is very different for active and quiet sets, as can be seen from the very different scale of the $y$-axis), whereas it strongly depends on this quantity when only resonant forces are taken into account. The fact that the embryos' mass at the end of our runs does not depend on the metallicity when thermal forces are included is confirmed in Fig.~\ref{fig:Median-Metalicidad}, which shows the median mass of the embryos (with $a>3$~au) reached at the end of the runs ($t=1$~Myr), as a function of metallicity, for the two kinds of runs (with resonant forces only and with resonant and thermal forces). We see that when only resonant forces are applied to the embryos, the median mass nearly scales with the metallicity, whereas it is nearly insensitive to this parameter when both thermal and resonant forces are applied. Note that the logarithmic $y$ axis on the plots of Fig.~\ref{fig:Median-Metalicidad} allows to gauge the relative mass dispersion from the height of the shaded region. This height is significantly smaller for the cases with all forces at high metallicity for the active sets (and comparable at low metallicity), while the opposite holds for quiet sets, for which the mass dispersion is much smaller with all forces at low metallicity.
The examination of Fig.~\ref{fig:synoptic} also brings up the following comments:
\begin{enumerate}
\item When embryos have sub-critical luminosity (i.e. $\Lambda < 0$, corresponding to thermal forces having a damping effect on the eccentricity and inclination), the median eccentricity of the embryos is smaller than that of embryos subjected to resonant forces only.
\item When embryos have a super-critical luminosity (i.e. $\Lambda > 0$), their median eccentricity is generally larger than that of embryos experiencing only the resonant forces. That median eccentricity is then a sizeable fraction of the gas disc's aspect ratio.
\item The dispersion of the eccentricity of the embryos experiencing thermal forces is generally smaller than that of the embryos experiencing only resonant forces, particularly at later time: thermal forces constrain more strongly the eccentricity (toward a value that depends on the luminosity) than resonant forces.
\item A similar observation can be done for the inclination, which always tends to show a smaller dispersion for embryos subjected to thermal forces.
\item The mass of the embryos experiencing thermal forces shows a dispersion marginally smaller or comparable to the dispersion of embryos experiencing resonant forces only.
\item Embryos at low metallicity have a sub-critical luminosity for a sizeable fraction of the duration of the runs, while they have straight away a super-critical luminosity at larger metallicity. Yet, since they have a similar mass growth for both values of the metallicity, their absolute luminosity is comparable. The different behaviour of the dimensionless luminosity arises because the critical luminosity $L_c$  at which thermal forces revert direction scales with the disc's thermal diffusivity \citep[][and Appendix~\ref{sec:thermal-force-shear}]{2019MNRAS.485.5035F}. At low metallicity, the gas is more transparent and radiative thermal diffusion is more efficient so it takes a larger luminosity to revert the sign of thermal forces. This was already seen in Fig.~\ref{fig:lcoverm} and in the maps of Fig.~\ref{fig:Mapa-Fiducial} for a single embryo, and will be further discussed in Section~\ref{sec:how-general}.
\end{enumerate}

\section{Discussion}
\label{sec:discussion}
The following results emerge from the previous sections, namely:
\begin{enumerate}
\item The inclusion of thermal forces, for our choice of parameters, in general speeds up accretion and renders the final median masses of the embryos nearly independent of the disc's metallicity, whereas such is not the case when only resonant forces are taken into account. This result is found both for the dynamically active and for the dynamically quiet setups. Thermal effects, for the disc parameters that were considered in this work, seem therefore to play a buffer effect against the variation of the metallicity. We discuss this effect in Section~\ref{sec:insens-discs-metall} below.
\item The relative mass dispersion in systems in which thermal forces are included
can be significantly smaller, or comparable to that in systems obtained with the resonant force only. In the systems in which this relative mass dispersion is small, the planets are more similar to one another. We mention the possible reasons for this behaviour in Section~\ref{sec:mass-disp-embry}.
\end{enumerate}

\subsection{Insensitivity to the disc's metallicity}
\label{sec:insens-discs-metall}
Our discussion firstly covers the results of our runs, then we address the question of the generality of our results.
\subsubsection{Analysis of the dynamically active and quiet runs}
\label{sec:active-case}
Let us first consider the runs of the dynamically active systems.
We see in the fourth row  of Fig.~\ref{fig:synoptic} how the ensemble averages of the median accretion efficiency $\varepsilon$ depends strongly on the metallicity when thermal forces are included, and how accretion is much more efficient at low metallicity.
\begin{figure}
\includegraphics[width=\columnwidth]{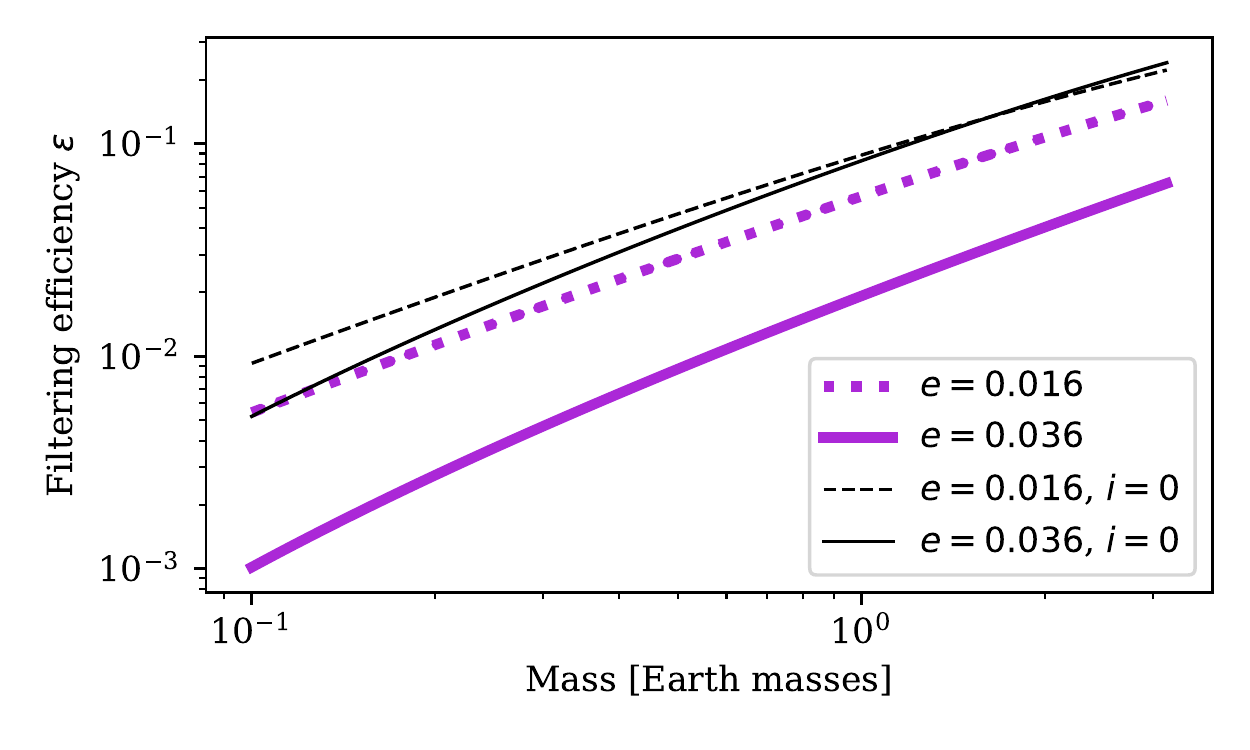}
  \caption{\label{fig:medfilt} Filtering efficiency as a function of the planet mass, for an eccentricity $e=0.036$, representative of the high metallicity case with thermal forces and for an eccentricity $e=0.016$, representative of the low metallicity case. The efficiency is calculated assuming either $i=e/2$ (thick curves) or $i=0$ (thin curves). There is a ratio $2.5-5$ between the accretion efficiency (depending on the mass) in favour of the case of low metallicity (hence low eccentricity), when the inclination is finite, whereas there is virtually no difference when the inclination is set to $0$.}
\end{figure}
The accretion efficiency is indeed a factor of several larger in the low metallicity case than in the high metallicity case when thermal forces are included, both in the active and quiet runs. It is this higher efficiency that makes up for the scarcity of solids: embryos accrete from a three times less abundant pebble flow with an efficiency nearly three times higher, resulting in similar median mass at the end of the simulations. What explains this increased efficiency at lower metallicity? The second row of Fig.~\ref{fig:synoptic} shows the (ensemble average of) the median eccentricity of the embryos as a function of time. When thermal forces are included, there is a significant difference between the low and high metallicity cases, whereas the difference is much milder in the purely resonant case. A similar statement holds for the inclination, shown in the third row. These large differences of the orbital elements account for the change in filtering efficiency, as can be seen on Fig.~\ref{fig:medfilt}, which shows the filtering efficiency as a function of mass, using the formulae from \citet{2018A&A...615A.138L} and \citet{2018A&A...615A.178O} with our disc parameters, for embryos with different eccentricities and inclinations. This figure shows four curves, obtained with two different values of the eccentricity: $e=0.036$ and $e= 0.016$ which are respectively the time-averaged eccentricity of the high metallicity case and that of the low metallicity case for the dynamically active runs. For each value of the eccentricity, we consider two possible choices for the inclination: either $e/2$ or $0$. For the first choice, we obtain a significant different accretion efficiency (embryos with the low values of eccentricity and inclination accrete several times more efficiently than those with the high values of these parameters). For the second choice, that of a zero inclination, the accretion efficiency is comparable for both values of the eccentricity. 
It therefore appears that the increased accretion efficiency in the case of low metallicity can be traced back to the inclinations of the embryos confined to lower values than in the high metallicity case. What is the reason for this lower inclination? The median mass has a similar time behaviour in the low and high metallicity cases, as we just saw, hence the accretion rates and  luminosities of the embryos ought to be similar in the two cases. Despite similar absolute values of the luminosity, embryos in the low and high metallicity cases have very different values of the \textit{dimensionless} luminosity $\Lambda$. We have indeed seen that the critical luminosity above which eccentricity and inclination are excited is smaller at higher metallicity (see Section~\ref{sec:evol-an-isol} and Fig.~\ref{fig:lcoverm}).
The bottom row of Fig.~\ref{fig:synoptic} shows the median value of the dimensionless luminosity. Both for active and quiet setups, embryos in the low metallicity case have a value of this parameter smaller than those with higher metallicity. For a given setup and at a given time, $\Lambda$ increases as the metallicity does, and consequently, the embryos are more dynamically excited at higher metallicity. From Fig.~\ref{fig:synoptic}, the median inclination in all our cases is at least $5\cdot 10^{-3}$, except at lower metallicity in the dynamically quiet cases. This value is also the aspect ratio of the disc of pebbles, which is $\sim\sqrt{\alpha_z/\tau_s}h$ \citep{1995Icar..114..237D}. When the embryo's inclination exceeds this value, they accrete only over a fraction of their orbit and their accretion efficiency scales as $i^{-1}$ \citep{2018A&A...615A.178O}: embryos in more metallic discs, which are more dynamically excited, have an accretion efficiency smaller than their peers in less metallic discs.

Such effects are absent from the simulations with purely resonant forces. The eccentricity and inclination of the embryos, in these circumstances, do not show a strong dichotomy between the low and high metallicity cases, and a similar statement holds for the accretion efficiency between both cases \textit{for a given planetary mass}. The strong disparity in the time behaviour of the accretion efficiencies of the resonant cases (fourth row of Fig.~\ref{fig:synoptic}) therefore simply reflects the fact that, at low metallicity, the embryos accrete from a much reduced pebble flow and grow more slowly.

\subsubsection{How general are our results?}
\label{sec:how-general}
Generalising our results to arbitrary cases is not straightforward, owing to the large parameter space involved and to the complex relationships between the orbital elements of the embryos and these parameters. We can nevertheless try and shed some light on the main reasons for the effect we have found by considering a simplified model in which a set of embryos has an equilibrium r.m.s. eccentricity resulting from the excitation by dynamical relaxation \citep[or viscous stirring][]{paplar2000,2002ApJ...581..666K}, the damping action of resonant wave excitation \citep{2004ApJ...602..388T} and the damping or excitation by thermal forces (depending on the sign of $\Lambda$).
The former is given by \citep{paplar2000}:
\begin{equation}
  \label{eq:vs1}
  \left.\frac{de}{dt}
  \right|_\text{vs}\simeq \frac 12e^{-3}\left(\frac{M_\text{ED}}{M}\right)\left(\frac{m}{M}\right)\Omega,
\end{equation}
where $M_{\text{ED}}$ is the mass of the embryos disc. We comment that, owing to the large uncertainty that come into the determination of the Coulomb logarithm, we do not write it in the equation above. The ratio of the largest to smallest impact parameters is of order of a few for the planet masses considered here and we assume the Coulomb logarithm to be of order unity. Eq.~\eqref{eq:vs1} is therefore given to within a factor of order unity. The damping by resonant wave excitation reads, with our notation \citep{2004ApJ...602..388T}:
\begin{equation}
  \label{eq:wd1}
  \left.\frac{de}{dt}
  \right|_\text{w}\simeq-\frac 18eh^{-4}\left(\frac{M_\text{GD}}{M}\right)\left(\frac mM\right)\Omega,
\end{equation}
where $M_\text{GD}=\pi\Sigma r^2$ is the (local) mass of the gaseous  disc, and where we have made use of Eq.~\eqref{eq:42} and specialised to the case $\gamma=1.4$. Finally, the action of thermal forces can be determined by noting that the values of eccentricities typically attained in our runs correspond to a radial excursion of the embryos larger than the thermal length scale $\lambda$ (see Appendix~\ref{sec:thermal}). In these circumstances, the thermal disturbance takes the form of a ``cometary trail'' \citep{2017A&A...606A.114C,2017arXiv170401931E}, and the time derivative of the eccentricity under the action of the thermal force is then given by \citep{2017arXiv170401931E,2023MNRAS.tmp..658C}:
\begin{equation}
  \label{eq:deth1}
  \left.\frac{de}{dt}
  \right|_\text{th}= 1.37\frac{F_\text{th}}{mr\Omega},
\end{equation}
where the expression of the thermal force, in that regime, is \citep{2020MNRAS.495.2063V}:
\begin{equation}
  \label{eq:dthct}
F_\text{th}=2\pi \left( \gamma-1 \right) \frac{ \left( Gm \right)^2\rho_0}{c_s^2r\Omega}\Lambda.
\end{equation}
As we shall see, it is desirable to improve slightly the above estimate to describe, in an approximate fashion, the decay of that force when the eccentricity is larger than the disc's aspect ratio, and the embryo is in a supersonic regime. In these circumstances, the force decays inversely proportionally to the speed squared \citep{2017MNRAS.465.3175M,2019MNRAS.483.4383V}. We therefore write:
\begin{equation}
  \label{eq:dthct2}
F_\text{th}=2\pi\left( \gamma-1 \right) \frac{\left( Gm \right)^2\rho_0}{c_s^2r\Omega}\frac{1}{1+\left( e/h \right)^2}\Lambda.
\end{equation}
Specifying again to the case $\gamma=1.4$, we can rewrite Eq.~\eqref{eq:dthct2} as:
\begin{equation}
  \left.\frac{de}{dt}
  \right|_\text{th}\simeq \frac 14 h^{-3}\left(\frac{M_\text{GD}}{M}\right)\frac{1}{1+(e/h)^2}\left(\frac mM\right)\Lambda\Omega.
\end{equation}
The equilibrium eccentricity is reached when:
\begin{equation}
  \label{eq:ecceq}
  \left.\frac{de}{dt}
  \right|_\text{vs}+\left.\frac{de}{dt}
  \right|_\text{w}+\left.\frac{de}{dt}
  \right|_\text{th}=0.
\end{equation}
Upon simplification, the equilibrium eccentricity is found to obey:
\begin{equation}
  e^{-3}\frac{M_\text{ED}}{M}-\frac 14 eh^{-4}\frac{M_\text{GD}}{M}+\frac 12h^{-3}\frac{1}{1+(e/h)^2}\frac{M_\text{GD}}{M}\Lambda=0.
\end{equation}
Introducing the `reduced eccentricity'' $\tilde e\equiv e/h$, the above equation can be simplified into:
\begin{equation}
  \label{eq:mainecceq}
  \xi^{-1}\tilde e^{-3}-\frac 14\tilde e+\frac 12\frac{\Lambda}{1+\tilde e^2}=0.
\end{equation}
where
\begin{equation}
  \label{eq:39}
\zeta\equiv \frac{M_\text{GD}}{M_\text{ED}}  
\end{equation}
is the ratio of the masses of the gas disc to the embryo disc.  For a given value of this ratio, there is therefore, a one-to-one relationship between the (reduced) equilibrium eccentricity and the dimensionless luminosity of the embryos. A few particular cases can be obtained from Eq.~\eqref{eq:mainecceq}. When $\Lambda=0$ (the thermal forces cancel out), the equilibrium eccentricity is of order $\tilde e = \sqrt{2}(M_\text{ED}/M_\text{GD})^{1/4}$. This trend is similar to that found by \citet{paplar2000}. When $M_\text{ED}/M_\text{GD}\rightarrow0 $, which corresponds to the case of an isolated embryo, Eq.~\eqref{eq:mainecceq} only admits a solution if $\Lambda>0$, and the eccentricity scales with the reduced luminosity as long as it is largely smaller than the disc's aspect ratio \citep{2022MNRAS.509.5622V}, i.e. as long as the reduced eccentricity is much smaller than one\footnote{The case of a vanishingly small eccentricity, which would apply to the case $\Lambda<0$, requires to use the formalism valid when the epicyclic excursion is smaller than the thermal length scale, rather than Eq.~\eqref{eq:deth1} \citep{2019MNRAS.485.5035F,2023MNRAS.tmp..658C}.}. For the systems of embryos that we have considered here, the gas to embryo mass ratio beyond the snow line varies from $\sim 300$ ($\sim 100$) for the quiet (active) case at $t=0$ to $\sim 30$ ($\sim 15$) at $t=1$~Myr.
\begin{figure}
 \includegraphics[width=\columnwidth]{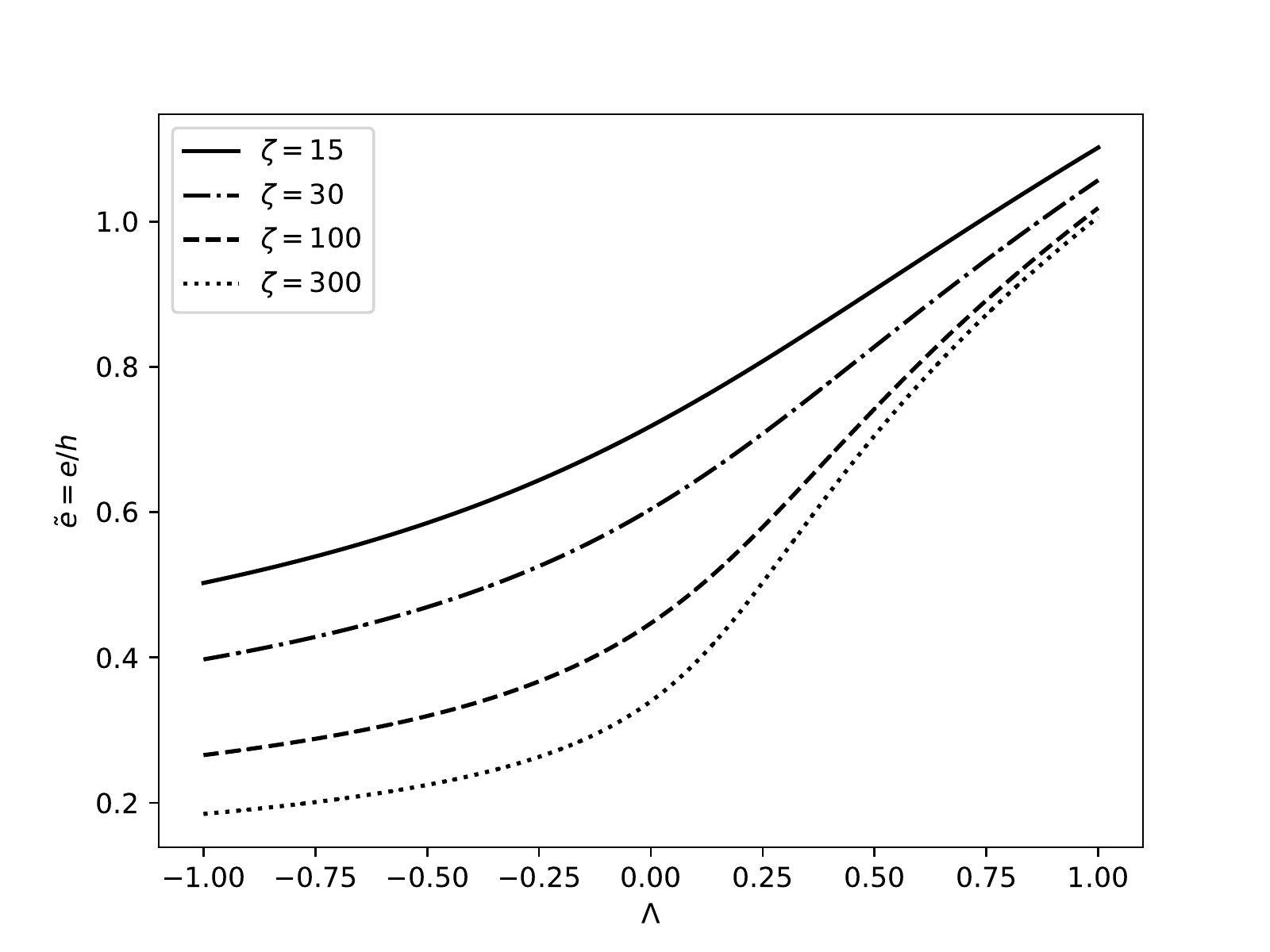}
 \caption{Reduced eccentricity $\tilde e$ versus dimensionless luminosity $\Lambda$, for different values of the gas to embryo mass ratio. When this ratio tends to infinity, one recovers the case of an isolated embryo, for which the eccentricity is zero for subcritical luminosity ($\Lambda < 0$) and is an increasing function of the luminosity when the latter is supercritical ($\Lambda >0$). If we had not introduced the factor in $(1+\tilde e^2)^{-1}$ in the expression of the thermal force (see Eq.~\ref{eq:dthct2}), there would be a steeper increase of the eccentricity for $\Lambda>0$, which would reach values significantly larger than the aspect ratio. }
 \label{fig:eccvslum}
\end{figure}
Fig.~\ref{fig:eccvslum} shows how the reduced eccentricity varies with the reduced luminosity for these values of the gas-to-embryo mass ratios. As could be expected, the eccentricity increases with the reduced luminosity, and the increase is more pronounced when the gaseous disc is much more massive than the embryos disc. In order to understand how the dimensionless luminosity relates to the metallicity, we write the expression of the dimensionless luminosity, using Eqs.~\eqref{eq:16}, \eqref{eq:18} and \eqref{eq:19} of Appendix~\ref{sec:thermal-force-shear}. Under the conditions that prevail beyond the snow line in our disc model, the embryos' masses are larger than the critical mass $M_c$, so that we make the approximations $\mathcal{C}_4\sim 4M_c/m$ and $\mathcal{C}_2\sim 2M_c/m$, which yield:
\begin{equation}
\label{eq:Llamb}
  \Lambda=\frac{\gamma Lc_s}{\pi \left( Gm \right)^2\rho_0}-\frac{2\chi c_s}{Gm}.  
\end{equation}
For a given planetary mass $m$ and luminosity $L$, the dimensionless luminosity, therefore, reads $\Lambda=a-b\chi$, and thus depends on the disc's metallicity exclusively through the $\chi$ factor. In each of our numerical experiments, the (median) masses of the embryos approximately follow similar laws as a function of time, hence the embryos have similar masses and luminosities at a given time, independently of the metallicity. In Eq.~\eqref{eq:Llamb}, the first term is therefore, approximately the same for the different metallicities. The difference in the $\Lambda$ behaviour between the different values of metallicity therefore stems exclusively from the differences in $\chi$, which scales with the inverse of the disc's metallicity. It is readily apparent from the plots of Fig.~\ref{fig:synoptic} that the curves of $\Lambda$ are similar indeed for different metallicities, except for a vertical shift: the higher the metallicity, the higher the value of $\Lambda$. By virtue of the graphs of Fig.~\ref{fig:eccvslum}, we infer that the higher the metallicity, the higher the r.m.s. eccentricity of the embryos (for similar masses and luminosities). Assuming for a moment that there is equipartition between horizontal and vertical motion, so that the r.m.s. inclination is half of the r.m.s. eccentricity, we conclude that the embryos also have larger inclinations, at a given time, in discs with higher metallicity, and therefore a smaller accretion efficiency, as discussed in Section~\ref{sec:active-case}.
While the relationship between the dimensionless luminosity and the eccentricity or inclination does not imply that the variation of the accretion efficiency exactly  compensates that of the mass flux of pebbles, it helps to understand why thermal forces tend to mitigate the consequences of the variation of metallicity. This discussion also gives some insight into the conditions for this effect to occur: it requires that (i) the gas disc is much more massive than the embryos' disc (the relation between the dimensionless luminosity and the eccentricity or inclination is steeper when this condition is fulfilled); (ii) the embryo masses are in the cut-off regime of thermal forces and (iii) the inclination of the embryos is larger than the aspect ratio of the pebbles' disc. None of these conditions is particularly stringent: the first one should be fulfilled over most of the life of the protoplanetary disc, the second one should be verified beyond the snow line when the embryo masses are in excess of a Martian mass \citep{2020arXiv200400874B}, especially as the disc ages \citep{2017MNRAS.465.3175M}. Finally, the third condition is supported by millimetre wavelength observations that show that the layer of millimetric dust is in general extremely thin \citep{2016ApJ...816...25P,2022A&A...668A.105D}, at least at distances to the star resolved by ALMA.

So far we have assumed that there is equipartition between the eccentricity and inclination, so that $\langle i^2\rangle^{1/2}=\langle e^2\rangle^{1/2}/2$. Yet we have mentioned in Section~\ref{sec:dynam-active-embry} that equipartition is not verified at low metallicity. We have not investigated this effect quantitatively, which can occur when the timescale for the damping of inclination becomes shorter than the timescale of dynamical relaxation. As the timescale for the evolution of orbital elements under the action of thermal forces is shorter at smaller metallicity (in the cut-off regime this timescale scales as $\chi^{-1/2}$),  this effect preferentially occurs at low metallicity, as can be seen in Fig.~\ref{fig:Resumen-Metalicidad}. It helps to further increase the accretion efficiency and make up for the paucity of pebbles when the metallicity is low.

The fact that the final mass of the embryos is nearly independent of the metallicity cannot hold for metallicities significantly smaller than the smallest value of our numerical experiments: for the lowest value of metallicity considered here, the eccentricity and inclinations have very low values at the beginning of the runs, and the embryos have a nearly maximal accretion efficiency. Should the metallicity be decreased further, the accretion efficiency would not increase significantly, but the inflow mass rate of pebbles would decrease, and so would the growth rate of the embryos. In a similar vein, at high metallicity, the embryos eccentricity is driven to values comparable to the disc's aspect ratio (and their inclination to half this value). Should the metallicity increase further, these orbital elements will not be driven to higher values: the accretion efficiency of pebbles should remain relatively constant, and the mass growth rate will then scale with the mass flux of pebbles.

\subsection{Mass dispersion of the embryos}
\label{sec:mass-disp-embry}
Fig.~\ref{fig:Median-Metalicidad} shows that the relative mass dispersion of the embryos can be quite small at low metallicity, when thermal forces are included. One reason for this can be found in the behaviour noticed in Section~\ref{sec:dynam-active-embry}: the dispersion of eccentricities and inclinations of embryos subjected to thermal forces is significantly smaller than that of embryos subjected only to resonant forces, since the driving timescale by thermal forces is much shorter than that due to the resonant interaction  (see Section~\ref{sec:evol-an-isol}). The accretion efficiency is a function of parameters intrinsic to the disc, and of the planet's mass, eccentricity and inclination. If these variables have little dispersion, so does the accretion efficiency. If two embryos have nearly similar accretion efficiency, they experience a similar mass growth (notwithstanding the fact that they are fed from a different mass flow of pebbles: the outer embryo sees a larger mass flow [see Appendix~\ref{sec:pebble-accretion}]).

This is particularly important at lower metallicity, where the damping timescale by thermal forces is the shortest. At higher metallicity, this timescale increases and the dispersion of eccentricity and inclination is larger.

At low metallicity, the strong driving of eccentricity and inclination by thermal forces therefore results, in addition to an enhanced accretion, to a more even growth rate among the different embryos.

\subsection{Comparison to previous work on thermal forces}
\label{sec:comp-prev-work}
Previous work on the impact of thermal forces on the growth and orbital evolution of low-mass protoplanets has so far been limited to the study of one embryo at a time. \citet{2022MNRAS.509.5622V} consider eccentric, low-mass embryos with zero inclination subjected to pebble accretion, and work out their equilibrium eccentricity  (the eccentricity that they acquire as a result of their own luminosity, and which provides them with an accretion rate that precisely yields this luminosity). That work only considers embryos with super-critical luminosity (as for an isolated embryo the equilibrium eccentricity of the sub-critical case is simply zero). Their disc has surface density and temperature that are power laws of the radius, like the disc considered in the present study. Counter to the present work that incorporates the dependence of the opacity on the density and temperature, they use a fixed opacity of $1$~cm$^2$.g$^{-1}$. The value of $\lambda/H$ that they report (the ratio of the thermal length scale to the pressure length scale) is comparable to the one obtained in the present analysis for $Z=Z_\odot$. While in the present analysis we consider only one value for the turbulent viscosity ($\alpha_z=10^{-4}$), they also consider smaller values ($\alpha_z=10^{-5}$ and $\alpha_z=0$). Also, while we consider here only  one value of the dimensionless stopping time ($\tau_s=10^{-2}$), they consider
five different values from $10^{-3}$ to $10^{-1}$. When $\alpha_z=10^{-4}$ and $\tau_s=10^{-2}$, and when the pebble mass rate matches the rate that we have adopted in the present study ($\dot M_\mathrm{peb}=10^{-4}\;M_\oplus.\mathrm{yr}^{-1}$), they find that the impact of thermal forces on the growth timescale of an embryo is nearly equal to that of an embryo maintained on a circular orbit, and that the eccentricity starts to grow when the embryo's mass exceeds $0.2-0.3\;M_\oplus$. This is broadly compatible with our results\footnote{Note that in our disc the dimensionless radial pressure gradient is $\eta\approx 2.5\cdot 10^{-3}$ instead of $2\cdot 10^{-3}$ in that considered by \citet{2022MNRAS.509.5622V}.} (see e.g. Fig.~\ref{fig:Mapa-Fiducial}). They do find a strong impact of thermal forces on the growth time at smaller turbulent viscosity (the growth time can then be halved if an embryo becomes eccentric). In that case, the embryo's luminosity is super-critical, and their results are a consequence, essentially, of the heating force. In our case, where we have a collection of interacting embryos with non-vanishing inclinations and a disc with a more significant turbulent viscosity, we find that the dominant effect is the maintenance of a low inclination of embryos with sub-critical luminosities, i.e. the most important factor to account for the marked difference between our two kinds of calculations (with thermal forces or without) is the action of the cold thermal force.

There has also been earlier work considering the impact of thermal forces on the growth and orbital evolution of isolated low-mass protoplanets, assuming those to be on circular orbits. While this assumption is valid as long as they have a sub-critical luminosity, it breaks down as soon as their luminosity exceeds the critical luminosity. This should occur much before the end of the calculations, in which planets with several Earth masses are usually obtained \citep{2019MNRAS.486.5690G, 2020arXiv200400874B,Guilera2021}. Earlier work, also, did not correctly evaluate thermal forces for masses in excess of the critical mass \citep[whose dependence was worked out by][see also Eq.~\ref{eq:8}]{2020MNRAS.495.2063V}.
In particular, an \textit{ad hoc}, manual cancellation of the force for planet masses in excess of the critical mass unsurprisingly leads to the outright, artificial cancellation of the effect of thermal forces, and can lead to the erroneous conclusion that these forces are unimportant \citep{2020arXiv200400874B}.

\subsection{Caveats, shortcomings and further remarks}
\label{sec:caveats-shortcomings}
We draw hereafter a list of the shortcomings and caveats of our  analysis:
\begin{enumerate}
\item We do not track the time evolution of the gaseous disc; 
\item We consider only perfect mergers, instead of a more complex variety of outcomes of collisions \citep[][and refs therein]{2021ApJ...921..163S};
\item We do not consider a possible change of size of the pebbles with time \citep{2014A&A...572A.107L} or at the snow line \citep[e.g.][]{2015Icar..258..418M}, and we consider only one Stokes number for the pebbles; 
\item Our simulations do not show the full evolution of the system up to the emergence of a planetary system devoid of gas. Rather, since we do not include the accretion of gas onto the rocky cores, our simulations are valid until a core reaches the pebble isolation mass (PIM).
\item We assume a constant mass flow of pebbles, instead of considering its decay over the $1$~Myr time frame of our calculations \citep{2014A&A...572A.107L}.
\item We consider simple power laws for the disc profiles, and resonant torque expressions that do not take into account all the effects at play in the coorbital region \citep{pbk11,2017MNRAS.471.4917J}. Rather, we introduce an \textit{ad hoc} migration trap at the snow line.
\item We do not consider the variations of the disc profiles under a change of metallicity. A change of metallicity in our setups only impacts the mass flow of pebbles ($\dot M_\mathrm{peb}\propto Z$) and the thermal diffusivity ($\chi\propto Z^{-1}$). Different laws may be obtained if one takes into account the change of the disc profiles when the metallicity changes. For instance, the temperature of viscously heated discs have a weak dependence on the opacity \citep[e.g.][]{2013A&A...549A.124B}. Accounting for this dependence leads to a more shallow dependence of the thermal diffusivity on the metallicity.
\item The initial mass and separation of our embryos is rather arbitrary, and so is the range of semi-major axis.
\item Our expression for the embryos' luminosity is simple and does not take into account the complexity of the accretion process, such as the ablation of pebbles in the envelope, the rain-out of a high metallicity vapour, etc. \citep{2018A&A...611A..65B}.
\item Similarly, the post-impact luminosity is evaluated in a very approximate fashion, and could be significantly higher if the merger acquires, during a transient period, a much larger size than that given by the conservation of volume \citep{2018JGRE..123..910L}.
\end{enumerate}
Points~(i) to~(viii) are not specific to the calculations with thermal forces: these simplifications affect systems subjected to thermal forces and systems subjected to resonant forces only. We do not expect any of these simplifying assumptions to be responsible for the considerable difference of outcomes between runs with all forces and runs with resonant forces only. 
As a consequence, even if we had performed a study including more realistic prescriptions regarding these points, we expect that thermal forces would still significantly boost accretion, especially at low metallicity, in a way similar to that discussed in Section~\ref{sec:insens-discs-metall}, and that they could induce a more orderly growth of the embryos, as discussed in Section~\ref{sec:mass-disp-embry}. Our results hinge essentially on the strong damping of eccentricity and inclination provided by the cold thermal force. This strong effect, in turn, arises because the thermal lengthscale $\lambda$ is much smaller, in general, than the pressure length scale $H$ \citep[][and Appendix~\ref{sec:thermal-force-shear}]{2017MNRAS.472.4204M}. Only in circumstances where these two length scales become comparable do we expect a minor impact of thermal forces.

Points~(ix) and~(x) correspond to processes that should have a minor impact on our findings. The phase of sub-critical luminosity is more crucial for the results presented here, and if the luminosity of the embryos were significantly smaller than that used in the present work (see Eq.~\ref{eq:2}), this would strengthen our conclusions. Among the effects that could decrease the luminosity with respect to the nominal value are the fact that part of the potential energy of the impinging pebbles is used to induce a change of state of the material accreted instead of being converted into heat, and the fact that the energy excess brought by the accretion of pebbles can be kept for a time comparable to or longer than the disc phase, for instance through atmospheric blanketing. Regarding point~(ix), we find the post-impact luminosity to have a negligible impact on the outcome of our calculations. Mergers are relatively rare in our simulations and the ``flash events'' they induce are short lived, so that they do not significantly contribute to our results. We have checked this with simulations in which we set the post-impact luminosity was set to zero, and did not find significant differences with our main set of calculations.

We also comment that the cold thermal torque acting on a low-mass planet in a disc with power law profiles of surface density and temperature induces a very fast inward migration \citep[][and this work]{2014MNRAS.440..683L,2017MNRAS.472.4204M}. This torque exacerbates the long-standing problem of too fast type~I migration by nearly one order of magnitude, and brings all ``cold'' objects to short period orbits on a short timescale. This seems to constitute a strong case against power law profiles in the inner parts of protoplanetary discs. Migration traps for objects dominated by the cold thermal torque are not of same nature as those abundantly studied in the literature, based on large spikes of the corotation torque at specific locations in the disc \citep{trap06,2010ApJ...715L..68L,2014A&A...569A..56C,2015MNRAS.452.1717L}. The sign of the thermal torque, for a planet on circular orbit, depends exclusively on the sign of the offset between orbit and corotation \citep{2021MNRAS.501...24C}. At locations where the disc is super-Keplerian, the cold thermal torque is positive, and it vanishes at a pressure maximum.

There is mounting evidence that protoplanetary discs are composed of multiple rings at all scales
\citep{2018ApJ...869L..46D}. Discs that seem smooth with standard image reconstruction techniques from ALMA's visibilities appear composed of rings when super-resolution techniques are used \citep{2022MNRAS.509.2780J}. In the same order of ideas, global simulations of protoplanetary discs that incorporate non-ideal MHD effects find a trend of discs to self-organise into axisymmetric structures reminiscent of the rings found with ALMA \citep{2017A&A...600A..75B,2020A&A...639A..95R}. Planet formation may well proceed within such rings \citep{2020A&A...638A...1M}. The dynamics of a system of embryos within a ring would present some differences with respect to the systems studied in the present work. The accretion of pebbles may not proceed from an inflow from the outer disc, but from the finite amount of pebbles present in the ring. The presence of a pressure maximum would naturally produce a trap for embryos with sub-critical luminosity. Note that while a pressure maximum constitutes a trap for sub-critical embryos, it acts as a repellent for embryos with super-critical luminosity (the heating torque is positive beyond the maximum, where the disc is sub-Keplerian, and negative inside). Embryos which would reach a super-critical luminosity would then be expelled from the ring, which would \textit{de facto} appear as a factory of similarly sized protoplanets. As the critical luminosity is reached earlier at higher metallicity, there would be an anti-correlation between the mass of the embryos released by the ring and the metallicity. While such processes definitely warrant a dedicated study, this discussion underlines that thermal torques are central to the interaction of embryos with the gas.

\section{Conclusions}
\label{sec:conclusions}
We have performed N-body simulations of the evolution of a set of planetary embryos subjected to pebble accretion, embedded in a gaseous disc with turbulent viscosity $\alpha_z=10^{-4}$. The back reaction of the disc on the embryos has been modelled in one of two ways: (i) using solely standard expressions for the exchange of angular momentum between the disc and the embryos based on resonant interactions and (ii) in addition to these, by including the thermal forces, arising from thermal diffusion in the gas surrounding the embryos. We consider two different kinds of initial distributions of embryos: one with $40$~Mars-mass bodies packed between~$3$ and $6$~au, and another one 
with $16$~Mars-mass bodies logarithmically spaced over the same range of semi-major axis, resulting in separations of $8$~mutual Hill radii. We vary the disc's metallicity from $0.5Z_\odot$ to $1.5Z_\odot$ (which in practice amounts to changing the mass rate of pebbles and the disc's thermal diffusivity). For each setup, we perform $30$ simulations with different random seeds in which the embryos are subjected to the resonant force from the disc, and $30$ other simulations in which they also experience the thermal force from the disc. We find that the inclusion of thermal forces has a considerable impact on the dynamics and growth of the embryos, which underlines the need to take these forces into account when considering embryos embedded in gas. In particular,
we find, for the two kinds of systems described above, that the median mass of the embryos at the end of the calculations is nearly independent of the disc's metallicity when thermal forces are included, whereas it nearly scales with the metallicity when they are not: thermal forces, for the disc model that we have considered, have a buffer effect against variations of the metallicity. This result arises from the strong damping by the thermal force of the eccentricity and inclination of low-luminosity embryos, especially when the metallicity is low, which favours a higher efficiency of pebble accretion and makes up for the scarcity of solids. At the higher metallicity, the effect of the disc forces tend to be attenuated by dynamical relaxation and the outcome of simulations with and without thermal forces are more similar.
We provide in the Appendix all the expressions required to implement thermal forces in an N-body code.

\section*{Acknowledgements}
S.C. acknowledges a scholarship from CONACyT, Mexico.
F.M. gratefully acknowledges support from grants UNAM-DGAPA-PASPA and UNAM-DGAPA-PAPIIT IG-101-620, and the University of Nice-Sophia Antipolis and the Laboratoire Lagrange at the Observatoire de la C\^ote d'Azur for hospitality. 

%%%%%%%%%%%%%%%%%%%%%%%%%%%%%%%%%%%%%%%%%%%%%%%%%%
\section*{Data Availability}
Our implementation of the code used in this study can be obtained from the corresponding author upon reasonable request.
 
%%%%%%%%%%%%%%%%%%%% REFERENCES %%%%%%%%%%%%%%%%%%

\bibliographystyle{mnras}
\bibliography{fmasset}

%%%%%%%%%%%%%%%%%%%%%%%%%%%%%%%%%%%%%%%%%%%%%%%%%%

%%%%%%%%%%%%%%%%% APPENDICES %%%%%%%%%%%%%%%%%%%%%

\appendix

\section{Embryo-gas interactions}
\label{sec:formulaciones}
We give hereafter the full detail of our implementation of the disc's force acting on the embryos. This force comes from two contributions: that arising from the resonant interaction with the disc, and that arising from thermal effects (which are non-resonant as heat diffusion is not wavelike). For each of these two main contributions, we calculate the force exerted on an embryo with small eccentricity and inclination\footnote{Namely smaller than the disc's aspect ratio for the resonant interaction, see \S~\ref{sec:resonant}, and smaller than the thermal length scale  divided by the embryo's semi-major axis for the thermal interaction, see \S~\ref{sec:thermal}.} to first order in these orbital parameters, and on an embryo with a large eccentricity or inclination (as approximated by a formula of dynamical friction), and we blend these forces so as to get a time behaviour of the eccentricity and inclination in agreement with published results over the full range of values taken by these orbital elements. This method is used twice, once for the resonant force and once for the thermal force.

In N-body simulations that implement the gravitational force from the gaseous disc, one must use prescriptions for the disc's force that reproduce the force exerted by the disc as closely as possible. \citet{ida2020} examined and discussed the various prescriptions for the disc's forces used in the literature, and proposed a generalised prescription that works both in the subsonic and supersonic regimes, based on formulae of dynamical friction. While the disc's force is not necessarily antiparallel to the velocity vector of the planet relative to the gas and exhibits a complex variation in magnitude and angle over an epicycle \citep{2004ApJ...602..388T}, the rationale of the formulation of \citet{ida2020} is that it produces the same variation of eccentricity, inclination and semi-major axis when time-averaged over an orbital period. In a similar manner, the direction and magnitude of the thermal force have a non-trivial behaviour over an epicycle \citep{2019MNRAS.485.5035F,2023MNRAS.tmp..658C}. In our implementation, we cannot forgo the determination of the embryos' orbital parameters at each timestep, however, since we need their eccentricity and inclination to evaluate the accretion rate of pebbles using the expressions of \citet{2018A&A...615A.138L} and \citet{2018A&A...615A.178O}. For this reason we use, for embryos with small eccentricities and inclinations, the exact expression of the force's magnitude and direction as a function of the orbital phase, as given respectively by \citet{tanaka2002} and \citet{2004ApJ...602..388T} for the resonant force, and by \citet{2017MNRAS.472.4204M} and \citet{2019MNRAS.485.5035F} for the thermal force. 

\begin{figure}
 \includegraphics[width=\columnwidth]{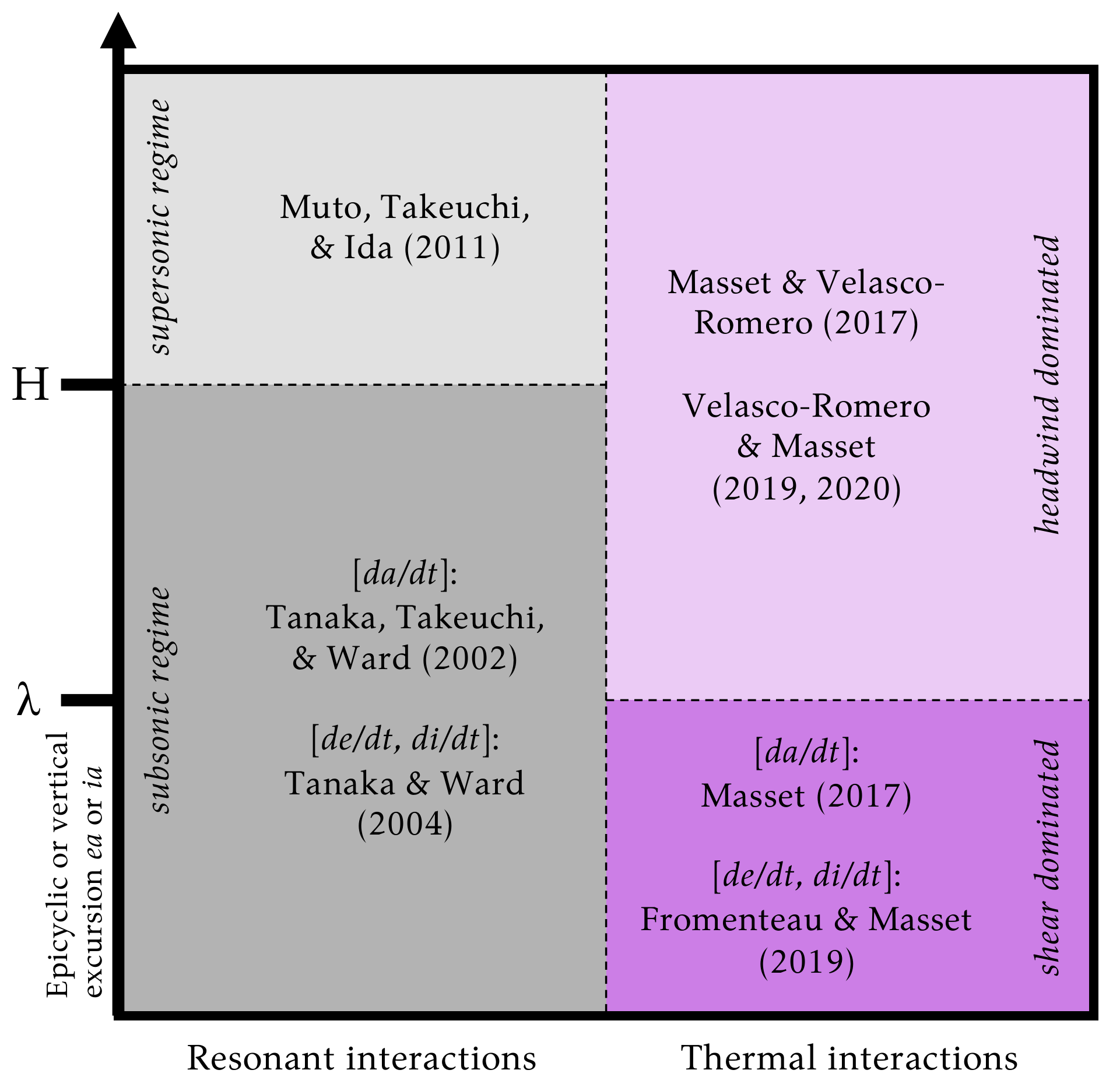}
 \caption{Graphical representation of the different regimes covered by our implementation of resonant and thermal forces. The epicyclic excursion $ea$  or vertical excursion $ia$ increase upwards. The left column shows the references we used to implement the resonant forces. A change of regime occurs when the excursion becomes $\sim H$.  The right column shows the references for the thermal forces. The change of regime, this time, occurs when the excursion is $\sim\lambda$.}
 \label{fig:tablita}
\end{figure}

We review the prescriptions for the resonant interactions in~\S\ref{sec:resonant} and for the thermal forces in~\S\ref{sec:thermal}. 
The different regimes and the references used for the different force expressions are summarised in Fig. \ref{fig:tablita}.

\subsection{Resonant interactions}
\label{sec:resonant}
In this section we present our evaluation of the resonant force in two limits: that of a small eccentricity and inclination (or limit of low Mach number), and that in which either the eccentricity or inclination is large (limit of large Mach number). We then work out a formulation that goes smoothly from one limit to the other.

\subsubsection{Limit of low Mach number}
\label{sec:limit-low-mach}
In this limit we work out the force with a first order expansion in eccentricity and inclination. The force is therefore the sum of a constant term, corresponding to the force exerted on a planet on a circular orbit, and an additional, time varying contribution arising from the radial and vertical excursions.

The force exerted on a planet on circular orbit has itself two components: that arising from the differential Lindblad torque, exerted by a wake excited by the embryo in the flow, and that arising from the corotation torque, which corresponds to an exchange of angular momentum between the embryo and the coorbital material. Together, these torques usually cause the embryo's inward migration \citep{tanaka2002}.

We use the expression of \citet{tanaka2002}, to which we add the \textit{ad hoc} trap at the snow line as discussed in Section~\ref{sec:spec-treatm-snowl}, to evaluate the force acting on the embryo if it is on a circular orbit: 
\begin{align}
    F^{[R]}_{0, \phi}= - \left\{ 1.364 + 0.541p + 5\exp\left[-\frac{(T-T_\mathrm{SL})^2}{\sigma_\mathrm{SL}^2}\right]\right\} q^{2} f_0h^2 \label{tanaka1},
\end{align}
where $p=-d\log\Sigma/d\log r$ is the slope of the surface density profile ($p=1/2$ in all our numerical experiments), and $f_0$ the reference force given by:
\begin{equation}
  \label{eq:5}
  f_0 = q^2\frac{\Sigma a^7\Omega_K^6}{c_s^4},
\end{equation}
where  $\Sigma$ and $c_s$ are the surface density and sound speed at the embryo's location, $q = m/M$ is the embryo to star mass ratio, $a$ is the semi-major axis of the embryo and $\Omega_K$ the Keplerian frequency at the embryo's semi-major axis.
In Eq.~\eqref{tanaka1} we set $T_\mathrm{SL}=170$~K and $\sigma_\mathrm{SL}=10$~K.
While there may also exist a radial component of the force exerted on an embryo on a circular orbit, we dismiss that component (as it contributes neither to the torque nor to the work exerted by the gas onto the embryo), and assume that the net resonant force exerted on a circular embryo is $\mathbf{F}^{[R]}=F^{[R]}_{0, \phi}\mathbf{e}_\phi$.

For embryos with low eccentricities ($ea \ll H$) and low inclinations ($i\ll h$), \citet{2004ApJ...602..388T} found that the first order terms of the force in eccentricity and inclination are:
\begin{align}
    F^{[R]}_{1,R}  = &  e f_0 \left( + 0.057 \cos f + 0.176 \sin f \right)  \\
    F^{[R]}_{1,\phi}  = &  e  f_0  \left( - 0.868 \cos f + 0.325 \sin f \right)  \label{tanaka2} \\
    F^{[R]}_{1,z}   = & i  f_0  \left[ -1.088 \cos \left( \omega + f \right)  -0.871 \sin \left( \omega + f \right) \right] 
\end{align}
where $f$ is the embryo's true anomaly, $\omega$ the argument of its periapsis, $e$ its eccentricity and $i$ its inclination.
The first order expansion in eccentricity and inclination of the net force exerted on the embryo in the low Mach number regime therefore reads:
\begin{equation}
  \label{eq:6}
  \mathbf F^{[R]}_{<}= (F^{[R]}_{0,\phi} +F^{[R]}_{1,\phi})\mathbf e_\phi+F^{[R]}_{1,R}  \mathbf e_R+F^{[R]}_{1,z} \mathbf e_z
\end{equation}

\subsubsection{Limit of large Mach number}
\label{sec:limit-large-mach}
Whenever the relative velocity $\mathbf V$ between the embryo and the ambient gas is larger than the speed of sound, we use the following expression:
\begin{align}
  \label{eq:7}
  \mathbf{F}^{[R]}_{>}= - \frac{\pi \Sigma G^{2} m^{2}}{ \epsilon c_s^2 }
  \frac{\mathcal{M}}{\mathcal{M}^3+3}\frac{ \mathbf{V}} {V},
\end{align}
where $\epsilon=0.6H$ is the softening length and $\mathcal{M}=V/c_s$ the embryo's Mach number with respect to the surrounding gas.
In the limit $\mathcal{M}\gg 1$, this expression tends asymptotically to that derived by \citet{2011ApJ...737...37M}, whereas it yields a force with magnitude comparable to that  given by \citet{1999ApJ...513..252O} in the subsonic regime. While this force is irrelevant in the subsonic regime, since the shear dominates the response and the force is given by Eq.~\eqref{eq:6}, it ensures that all the force formulae that we use in our final blending expression, given in the following section, are continuous functions of the embryo's velocity.

\subsubsection{Net resonant force for an arbitrary velocity of the embryo}
\label{sec:net-resonant-force}
Whether a first order expansion of the force in eccentricity and inclination is appropriate to calculate the force on an embryo or whether a calculation of dynamical friction is appropriate depends on the Mach number of the embryo \citep{2002A&A...388..615P}. We therefore seek an expression of the net force for an arbitrary Mach number of the embryo of the form:
\begin{equation}
  \label{eq:10}
  \mathbf{F}^{[R]}_0=(1-\alpha) \mathbf{F}^{[R]}_{<}+\alpha\mathbf{F}^{[R]}_{>}, 
\end{equation}
where $\alpha$ is a function of the Mach number, such that
\begin{equation}
  \label{eq:48}
  \left\{
    \begin{array}{lcl}
      \lim_{\mathcal{M}\rightarrow 0}\alpha=0&\mbox{and} &\lim_{\mathcal{M}\rightarrow \infty}\alpha=1   \\
      \lim_{\mathcal{M}\rightarrow 0}\frac{\alpha|\mathbf{F}^{[R]}_{>}|}{|\mathbf{F}^{[R]}_{<}|}=0
      &\mbox{and} &\lim_{\mathcal{M}\rightarrow \infty}\frac{(1-\alpha)|\mathbf{F}^{[R]}_{<}|}{|\mathbf{F}^{[R]}_{>}|}=0
    \end{array}
    \right.
\end{equation}
where the last two conditions entail that at small (large) Mach number, the force only consists of $\mathbf{F}^{[R]}_{<}$ ($\mathbf{F}^{[R]}_{>}$).
We find that the expression:
\begin{equation}
  \label{eq:11}
  \alpha(\mathcal{M}) = \frac{\mathcal{M}^3}{0.42+\mathcal{M}^3},
\end{equation}
which satisfies the above conditions, reproduces accurately the behaviour of $\dot e/e$ given by \citet{2007A&A...473..329C} for $e$ ranging from values small compared to $h$ to several $h$.

Finally, in order to account for the gas rarefaction at large distance from the midplane, we adopt the following, final expression for the force arising from the resonant interaction with the disc:
\begin{equation}
  \label{eq:12}
  \mathbf{F}^{[R]}=\exp(-z^2/2H^2) \mathbf{F}^{[R]}_0.
\end{equation}
For an embryo with low Mach number, which is always at a distance $|z|\ll H$ from the midplane, the correction factor in Eq.~\eqref{eq:12} is nearly unity, whereas for an embryo with a large Mach number, it simply amounts to evaluating the dynamical friction using the local density. This approximation is justified as the disc's response is nearly local for a supersonic embryo \citep{2002A&A...388..615P}. Adopting this prescription gives a time behaviour of the inclination in agreement with that given by \citet{2007A&A...473..329C}.

\subsection{Thermal interactions}\label{sec:thermal}
The calculation of the net thermal force on an embryo with arbitrary velocity with respect to the gas is similar to that of the resonant force. We evaluate the force to first order in eccentricity and inclination, suitable for an embryo with a low velocity, then the force given by a calculation of dynamical friction, suitable for an embryo with a larger velocity, and we take a combination of both with blending coefficients that depend on the embryo's velocity with respect to the gas. The typical radial excursion that corresponds to the transition between the two regimes is not the pressure length scale (corresponding to a Mach number of order unity) as is the case for the resonant force (see~\S~\ref{sec:net-resonant-force}) but rather the thermal length scale $\lambda$ in a sheared medium:
\begin{equation}
  \label{eq:13}
  \lambda=\sqrt{\frac{\chi}{(3/2)\Omega\gamma}},
  \end{equation}
  where $\chi$ is the disc's thermal diffusivity, given by Eq.~\eqref{eq:14}.
  For the disc we consider, this length scale is significantly smaller than that of pressure, so that the transition between both regimes occurs for smaller values of eccentricity or inclination (see Fig.~\ref{fig:tablita}). We refer to the case where the radial and vertical excursions are smaller than the thermal length scale as the shear dominated regime, whereas the case in which the radial or vertical excursion is larger than the thermal length scale is referred to as the headwind dominated regime.

\subsubsection{Thermal force in the shear dominated regime}  
\label{sec:thermal-force-shear}
As in \S\ref{sec:limit-low-mach}, we write the force to first order in eccentricity and inclination. The zeroth order term consists of the azimuthal component of the force exerted on a planet on a circular orbit (as in \S\ref{sec:limit-low-mach} we discard the effect of the radial component). This force reads, for a low mass planet:
\begin{equation}
  \label{eq:8}
F^{[T]}_{0,\phi}=1.61(\mathcal{C}_4\ell-\mathcal{C}_2)f_0K\eta h^2,
\end{equation}
where
\begin{equation}
  \label{eq:38}
  K = \gamma(\gamma-1)\frac H\lambda
\end{equation}
and $\eta$ is a dimensionless number of order unity that quantifies the distance between the orbit and corotation, and which has value $\eta=0.83$ in our disc \citep{2017MNRAS.472.4204M}.
In Eq.~\eqref{eq:8}, the variable $\ell$ is defined as:
\begin{equation}
  \label{eq:15}
  \ell = \frac{L}{L_c},
\end{equation}
where $L$ is the embryo's luminosity and $L_c$ is the critical luminosity near which thermal forces change sign:
\begin{equation}
  \label{eq:16}
  L_c=4\pi GM_p\chi\rho_0/\gamma.
\end{equation}
The numerical coefficients $\mathcal{C}_4$ and $\mathcal{C}_2$ account for the decay of thermal effects for an embryo mass past the critical mass $M_c$ given by \citet{2020MNRAS.495.2063V}:
\begin{equation}
  \label{eq:17}
  M_c=\frac{\chi c_s}{G}.
\end{equation}
While these coefficients were obtained through an empirical fit in the context of dynamical friction (i.e. that of a perturber moving through a homogeneous medium at rest), the results obtained by \citet{2021MNRAS.501...24C} for a planet on a circular orbit, and those obtained by \citet{2017arXiv170401931E} for an eccentric or inclined planet are compatible with a similar law for the decay of the thermal force. We therefore use the values worked out by \citet{2020MNRAS.495.2063V}, regardless of the regime (i.e. either shear or headwind dominated). They read:
\begin{equation}
  \label{eq:18}
    \mathcal{C}_2 \equiv \frac{2 M_{c}}{m+ 2M_{c}}; \hspace{0.5cm} \mathcal{C}_4 \equiv \frac{4 M_{c}}{m+ 4M_{c}}
\end{equation}
in all instances of thermal forces in the present work. For the sake of conciseness we denote with $\Lambda$ the pre-factor in Eq.~\eqref{eq:8}:
\begin{equation}
  \label{eq:19}
  \Lambda\equiv \mathcal{C}_4\ell-\mathcal{C}_2.
\end{equation}
Therefore $\Lambda$ is a dimensionless variable which depends on the embryo's luminosity. For a non-luminous embryo, it has a negative value that tends to $-1$ in the low-mass limit. In that same limit, when $L=L_c$, $\Lambda\rightarrow 0$: thermal effects cancel out. Finally, for a luminosity larger than $L_c$, it has a positive value. When the embryo's mass is largely sub-critical, we have $\Lambda\approx\ell -1$.

The first order terms in eccentricity and inclination are given by \citet{2019MNRAS.485.5035F}. They read:

\begin{align}
  \label{eq:20}
    F^{[T]}_{1,R}  = & \sqrt{2/\pi}e   \Lambda f_0 K \left( - 0.507 \cos f + 1.440 \sin f \right) 
  \\
  \label{eq:21}
    F^{[T]}_{1, \phi}  =& \sqrt{2/\pi}e  \Lambda f_0 K \left(  0.737  \cos f + 0.212 \sin f \right) 
  \\
  \label{eq:22}
    F^{[T]}_{1, z}  = & \sqrt{2/\pi}i   \Lambda f_0 K \left[ 1.160  \cos \left( \omega + f \right)  + 0.646 \sin \left( \omega + f \right) \right].
\end{align}
Note that a factor $(\ell-1)$ features in the analysis of \citet{2019MNRAS.485.5035F}, but we use $\Lambda$ instead, as mentioned above, to account for the decay of thermal effects at higher mass.

The net thermal force exerted on the embryo with low eccentricity and inclination therefore reads:
\begin{equation}
  \label{eq:66}
  \mathbf F^{[T]}_<= (F^{[T]}_{0, \phi}+F^{[T]}_{1, \phi})\mathbf e_\phi+F^{[T]}_{1, R } \mathbf e_R+F^{[T]}_{1, z}\mathbf e_z.
\end{equation}

\subsubsection{Thermal force in the headwind dominated regime}
\label{sec:headwind-regime}
This regime was analysed by \citet{2017MNRAS.465.3175M}, \citet{2019MNRAS.483.4383V} and \citet{2020MNRAS.495.2063V} who obtained that the force is given by 
\begin{align}
\mathbf{F}^{[T]}_>= \Lambda  \frac{2  \pi \left( \gamma-1 \right) \rho_{0} G^{2} m^{2}}{c_{s}^{2}} g \left( \mathcal{M} \right) \frac{  \mathbf{V}}{ \lvert  \mathbf{V} \rvert } 
\end{align}
where $g(\mathcal{M})$ is a function that tends to $1$ at low Mach number, and that exhibits a decay in $\mathcal{M}^{-2}$ in the supersonic regime.
For the behaviour of $g(\mathcal{M})$ in the supersonic regime, we take the expression given by \citet{2017MNRAS.465.3175M}, so that
\begin{equation}
  \label{eq:23}
  g(\mathcal{M}) = \xi/\mathcal{M}^2 \mbox{~~if $\mathcal{M} \ge 1$},
\end{equation}
where
\begin{equation}
  \label{eq:24}
  \xi = \log \left\lbrace \exp \left[ -1.96 - \log \left( \frac{ r_{\ast}  V}{4 \chi} \right)  \right] + 1 \right\rbrace.  
\end{equation}
In Eq.~\eqref{eq:24}, $r_{\ast}$ stands for the truncation radius for the force calculation, for which we take
\begin{equation}
  \label{eq:25}
  r_{\ast} = \frac{G m }{ \textup{max} \left(  V ,c_{s} \right)^{2} }.
\end{equation}
Eq.~\eqref{eq:24} is an approximation to the function $f$ of \citet{2017MNRAS.465.3175M} that ensures it decays to $0$ when its argument is larger than $1$.

There is no analytic expression of the thermal force available at arbitrary Mach number in the subsonic regime. We make use of the fact that the force is continuous at the transonic point, and has a nearly vanishing derivative in the Mach number for $\mathcal{M}\rightarrow 0$ \citep{2019MNRAS.483.4383V}, to write:
\begin{equation}
  \label{eq:26}
  g(\mathcal{M})=1+(\xi-1)\mathcal{M}^2 \mbox{~~if $\mathcal{M} < 1$}.
\end{equation}

\subsubsection{Net thermal force for an arbitrary velocity of the embryo}
\label{sec:net-thermal-force}
We use a prescription similar to that of \S~\ref{sec:net-resonant-force} and write
\begin{equation}
  \label{eq:27}
    \mathbf{F}^{[T]}_0=(1-\beta) \mathbf{F}^{[T]}_<+\beta\mathbf{F}^{[T]}_>.
  \end{equation}
  Here $\beta\in[0,1]$ is a blending coefficient that we evaluate by comparing the response timescales of the two versions of the force, in order to assess which one is dominant. In the headwind dominated regime, the response timescale is \citep{2017MNRAS.465.3175M}:
  \begin{equation}
    \label{eq:28}
 \tau_\mathrm{hw} = \frac{\chi}{\gamma^{2} V^{2}},
\end{equation}
whereas in the shear dominated regime, it is the shear timescale:
\begin{equation}
  \label{eq:29}
  \tau_\mathrm{sh}=\frac{2}{3\Omega}.
\end{equation}
The ratio of these two timescales:
\begin{equation}
  \label{eq:30}
  {\cal T}=\frac{\tau_\mathrm{sh}}{\tau_\mathrm{hw}}
\end{equation}
can be recast as ${\cal T}=(4/9)\gamma(V/\Omega\lambda)^2$, so that it scales with the squared ratio of the typical elongation of the planet to the thermal length scale. When ${\cal T}\gg 1$, the embryo is in the headwind dominated regime, while it is in the shear dominated regime if ${\cal T} \ll 1$. The $\beta$ function must therefore tend to one when ${\cal T}\gg 1$, and to zero when ${\cal T} \ll 1$. We have adopted:
\begin{equation}
  \label{eq:31}
  \beta({\cal T}) = \frac{{\cal T}^2}{1+{\cal T}^2},
\end{equation}
which satisfies conditions similar to those of Eq.~\eqref{eq:48}.
This prescription yields results that match satisfactorily the behaviour reported in numerical simulations, as can be seen in Section~\ref{sec:evol-an-isol}.

Taking into account the gas rarefaction at higher altitude is done differently than for the resonant case.  Heating forces are independent of the background density (up to a certain luminosity), as they arise from a deficit of density that depends on the luminosity of the perturber, the sound speed and the thermal diffusivity, but not on the unperturbed density. We work out the distance to the embryo at which the perturbation of density has same absolute value as the background density, and use that distance to evaluate the resulting truncation of the force, in a fashion similar to that done by \citet{2020MNRAS.495.2063V}.
In the vicinity of the perturber, the perturbation of density $\delta\rho$ has the dependence:
\begin{equation}
  \label{eq:32}
  \delta\rho\approx-\frac{\gamma(\gamma-1)L}{4\pi c_s^2\chi r},
\end{equation}
where $r$ is the distance to the perturber, assumed to be smaller than the characteristic length scales arising from the flow's geometry ($\lambda$ in the shear dominated regime or the size of the hot trail $\chi/\gamma V$ in the headwind dominated regime). The truncation distance is therefore:
\begin{equation}
  \label{eq:33}
  r_\mathrm{trunc}=\frac{\gamma(\gamma-1)L}{4\pi  c_s^2\rho_0}\exp(z^2/2H^2),
\end{equation}
and the reduction factor applied to the force is:
\begin{equation}
  \label{eq:34}
  z_\mathrm{cut} = \frac{2r'+\exp(-2r')-1}{2r'^2},
\end{equation}
where $r'=\gamma r_\mathrm{trunc} V/(2\chi)$.
\textit{In fine}, the net thermal force applied to each embryo is:
\begin{equation}
  \label{eq:35}
  \mathbf{F}^{[T]}=z_\mathrm{cut}\mathbf{F}^{[T]}_0.
\end{equation}
We also mention that in an early set of test runs we applied for the thermal force the same cut off procedure as for the resonant force (Eq.~\ref{eq:12}) and virtually found no difference with the results presented here.

\subsection{Luminosity of the embryos}
\label{sec:luminosity-embryos}
Evaluating the thermal forces requires to know the luminosity of the embryos. This quantity, in our implementation, arises from two processes:
\begin{enumerate}
\item the accretion of pebbles;
\item the conversion into heat of the kinetic energy after impacts (mergers) of two embryos.
\end{enumerate}
We give hereafter the details of the implementation of each of these processes.
\subsubsection{Pebble accretion}
\label{sec:pebble-accretion}
At each time step, we evaluate the accretion rate of pebbles on each embryo following exactly\footnote{We downloaded the library available at \url{https://staff.fnwi.uva.nl/c.w.ormel/software.html}, translated it to~C (the language used by REBOUND) and checked that our implementation yields same results to machine accuracy for a large set of arbitrary parameters.} the recipe given by \citet{2018A&A...615A.138L} and \citet{2018A&A...615A.178O}. This recipe provides for each embryo $i$ an efficiency $\varepsilon_i\in[0,1]$ of accretion of the inward pebble mass flux, as a function of the Stokes number $\tau_s$ of the pebbles, the embryo to star mass ratio, the turbulence parameter $\alpha_z$ of the disc, the gas disc's aspect ratio, the distance between the semi-major axis and corotation arising from the pressure support to the rotational equilibrium of the disc, the physical radius of  the embryo, its eccentricity and its inclination. The inward pebble mass flux is determined as follows.
At each time step, we sort the embryos by decreasing distance to the star. Denoting $D$ the map corresponding to this sort ($\forall j,k \in [0,n-1]$,  if $j<k$ then $r_{D(j)} > r_{D(k)}$), we calculate the global filtering $(\mathcal{F}_i)_{i\in[0,n-1]}$ of the mass influx of pebbles as:
\begin{equation}
  \label{eq:36}
  \mathcal{F}_i=\prod_{j=0}^{D^{-1}(i)-1}1-\varepsilon_{D(j)},
\end{equation}
where the product is meant to be $1$ if $D^{-1}(i)=0$ (i.e. if the $i$-\textit{th} embryo is the outermost one, there is no outer embryo that reduces the incoming pebble flux). The accretion rates are then determined as:
\begin{equation}
  \label{eq:37}
  \dot m_i=\dot M_\mathrm{peb}\mathcal{F}_i\varepsilon_i.
\end{equation}
In this equation, $\dot M_\mathrm{peb}\mathcal{F}_i$ is the mass flow of pebbles that reach the orbit of embryo~$i$.
This accretion rate is in turn used to:
\begin{enumerate}
\item evaluate the embryos' luminosities using Eq.~\eqref{eq:2} and
  \item update the masses of the embryos.
\end{enumerate}

\subsubsection{Post-impact luminosity}
\label{sec:post-impact-lumin}
To each embryo $i$ we associate an extra variable $\Delta E_i$ corresponding to an excess of internal energy. Each such variable is initialised to $0$ at the start of the simulation.
After a merger event, the energy excess of the merger is incremented by the difference of kinetic energies prior to and after the even. This energy excess is then radiated away with the luminosity:
\begin{equation}
  \label{eq:40}
  L_{\mathrm{PIL},i} = 4\pi R_i^2\sigma T_\mathrm{sil}^4,
\end{equation}
where $T_\mathrm{sil}=2,300$~K is the vaporisation temperature of silicates, as long as $\Delta E_i>0$. The post-impact luminosity $L_{\mathrm{PIL},i}$ is added to that arising from pebble accretion.

\subsection{Summary of the simulation parameters}
\label{sec:summ-simul-param}
We have used the values shown in Table \ref{tab:valores} to perform our simulations.

\begin{table}
\begin{center}
\begin{tabular}{| l |c | c | }
\hline
Quantity & Notation & Value \\
\hline
Aspect ratio & $h$ & 0.045 \\ 
Distance unit & $r_{0}$ & 1 au \\
Surface density at distance unit & $\Sigma_{0}$ & 400 g cm$^{-2}$ \\ 
Adiabatic index & $\gamma$ & 1.4 \\
Slope of surface density & $p$ & 0.5 \\
Turbulence parameter & $\alpha_z$ & $10^{-4}$ \\
Pebbles' Stokes number & $\tau_{s}$ & 0.01 \\
Pebble mass flux & $\dot M_\mathrm{peb}$ & $10^{-4}(Z/Z_\odot)$ M$_{\oplus}$ yr$^{-1}$ \\
\hline
\end{tabular}
\caption{Values of some constants used in the numerical experiments.}
\label{tab:valores}
\end{center}
\end{table}

% Don't change these lines
\bsp	% typesetting comment
\label{lastpage}
\end{document}